\documentclass[structabstract]{aa}

\usepackage{natbib}			
\usepackage[varg]{txfonts}		
\usepackage{nameref}
\begin{document}
  \titlerunning{From planet building blocks to rocky planets}
  \title{Linking the primordial composition of planet building disks to the present-day composition of rocky exoplanets}

  \author{V.~Adibekyan\inst{1,2} \and
          M.~Deal\inst{3} \and
          C.~Dorn\inst{4} \and
          I.~Dittrich\inst{4} \and
          B.~M.~T.~B.~Soares\inst{1,2} \and
          S.~G.~Sousa\inst{1} \and
          N.~C.~Santos \inst{1,2} \and
          B.~Bitsch \inst{5} \and
          C.~Mordasini\inst{6} \and
          S.~C.~C.~Barros\inst{1,2} \and
          D.~Bossini\inst{7,8} \and
          T.~L.~Campante\inst{1,2} \and
          E.~Delgado~Mena \inst{1} \and
          O.~D.~S.~Demangeon\inst{1} \and
          P.~Figueira\inst{9, 1} \and
          N.~Moedas\inst{1,2} \and
          Zh.~Martirosyan\inst{10} \and
          G.~Israelian\inst{11,12} \and 
          A.~A.~Hakobyan\inst{13}
 }

  \institute{
  	  Instituto de Astrof\'isica e Ci\^encias do Espa\c{c}o, Universidade do Porto, CAUP, Rua das Estrelas, 4150-762 Porto, Portugal \\
  	  \email{vadibekyan@astro.up.pt} \and
  	  Departamento de F\'{\i}sica e Astronomia, Faculdade de Ci\^encias, Universidade do Porto, Rua do Campo  Alegre, 4169-007 Porto, Portugal \and
  	  LUPM, Universit\'{e} de Montpellier, CNRS, Place Eug\`{e}ne Bataillon, 34095 Montpellier, France \and
  	  ETH Zurich, Institute for Particle Physics and Astrophysics, Wolfgang-Pauli-Strasse 27, CH-8093 Zurich Switzerland \and
  	  Department of Physics, University College Cork, Cork, Ireland \and
     Space Research and Planetary Sciences, Physics Institute, University of Bern, Gesellschaftsstrasse 6, 3012 Bern, Switzerland \and
  	  Dipartimento di Fisica e Astronomia Galileo Galilei, Università di Padova, Vicolo dell’Osservatorio 3, I-35122 Padova, Italy \and
  	  Osservatorio Astronomico di Padova -- INAF, Vicolo dell’Osservatorio 5, I-35122 Padova, Italy \and
            Observatoire Astronomique de l’Université de Genève, Chemin Pegasi 51b, 1290 Versoix, Switzerland \and
  	  Department of Astrophysics and General Physics, Yerevan State University, A. Manukyan str. 1, 0025 Yerevan, Armenia \and
  	  Instituto de Astrof\'{i}sica de Canarias, E-38205 La Laguna, Tenerife, Spain \and           
  	  Departamento de Astrof\`{i}sica, Universidad de La Laguna, E-38206 La Laguna, Tenerife, Spain \and
  	  Center for Cosmology and Astrophysics, Alikhanian National Science Laboratory, 2 Alikhanian Brothers Str., 0036 Yerevan, Armenia
  	  }

  \date{Received date / Accepted date }
  \abstract
    {The composition of rocky planets is strongly driven by the primordial materials in the protoplanetary disk, which can be inferred from the abundances of the host star. Understanding this compositional link is crucial for characterizing exoplanets.}
    {We aim to investigate the relationship between the compositions of low-mass planets and their host stars.}
    {We determined the primordial compositions of host stars using high-precision present-day stellar abundances and stellar evolutionary models. These primordial abundances were then input into a stoichiometric model to estimate the composition of planet-building blocks. Additionally, we employed a three-component planetary interior model (core, mantle, water in different phases) to estimate planetary compositions based only on their radius and mass.}
    {We found that although stellar abundances vary over time, relevant abundance ratios like Fe/Mg remain relatively constant during the main sequence evolution for low temperature stars. A strong correlation is found between the iron-to-silicate mass fraction of protoplanetary disks and planets, while no significant correlation was observed for water mass fractions. The Fe/Mg ratio varies significantly between planets and their stars, indicating substantial disk-driven compositional diversity, and this ratio also correlates with planetary radius.}
    {While stellar abundances, as a proxy of the composition of protoplanetary disk, provide a baseline for planetary composition, significant deviations arise due to complex disk processes, challenging the assumption of a direct, one-to-one elemental relationship between stars and their planets.}
  


  \maketitle
   
\section{Introduction}					\label{sec:intro}

With the rapid advance in exoplanet science, it is now possible to characterize not only the fundamental parameters (mass and radius) of planets, but also their interior structure and bulk composition \citep[e.g.][]{Dorn-18, Nettelmann-21, Adibekyan-21, Adibekyan-21b, Armstrong-20, Helled-22, Boley-23}. These parameters are important for determining the habitability conditions of terrestrial planets \citep[e.g.][]{Shahar-19, Unterborn-22}, and understanding the bulk composition is essential for unraveling the origins of planetary diversity \citep[e.g.][]{Owen-17, Jin-18, Bitsch-19, Venturini-20}.

Information about the mass and radius of exoplanets alone is insufficient to accurately characterize the interior composition of rocky exoplanets \citep[e.g.][]{valencia_detailed_2007, Dorn-15, Unterborn-23}. To break this degeneracy, the abundances of the main rocky-forming elements (Mg, Si, and Fe) in the host star are often used. Stellar atmospheres offer the only observational window into the remnants of the planet-forming disk's composition, as these disks, with lifetimes typically lasting only a few million years, have long dissipated around most discovered planet-hosting stars.

When modeling the interiors of rocky planets, a one-to-one correlation between the composition of the planet and that of its host star is typically assumed \citep[e.g.][]{Thiabaud-15, Dorn-15, Unterborn-16}. This assumption, to a first-order, appears to be valid for Earth \citep{McDonough-03}, Venus \citep{Aitta-12}, Mars \citep{Khan-18}, and the Sun \citep[see][]{Dorn-15, Unterborn-16}. It may also be valid for some exoplanets \citep{Santos-15, Schulze-21}, but not necessarily for others \citep{Bonomo-19, Plotnykov-20, Brinkman-23, Mah-23}.

Assuming a one-to-one compositional correlation is a reasonable starting point, but it breaks down even within our own Solar System, in the case of Mercury \citep{morgan_chemical_1980}. \citet{Adibekyan-21}\footnote{See also \citet{Plotnykov-20, Schulze-21, Wang-22, Liu-23, Spaargaren-23, Unterborn-23, Schulze-24, Brinkman-24}.} conducted a statistical comparison of the iron-to-silicate mass fraction in low-mass exoplanets and their host stars, concluding that while there is a correlation between the composition of rocky planets and their host stars, the relationship is not one-to-one. The analysis focused on 22 low-mass (M $<$ 10 $M_{\mathrm{\oplus}}$) and small-sized (R $<$ 2 $R_{\mathrm{\oplus}}$) planets with mass and radius measurements precise to within 30\%, all orbiting F, G, and K-type stars. However, the study by \citet{Adibekyan-21} has potential limitations: (i) The planets' interiors were modeled under the assumption that they consist only of a core and mantle, with no volatiles included, and (ii) the present-day stellar abundances were used as a proxy for the primordial protoplanetary disk composition. Some astrophysical processes, such as atomic diffusion, can alter a star's surface composition as it evolves \citep{deal18}, meaning that the present-day stellar abundances could differ significantly from their primordial values.

In this work, we study the star-planet compositional link by addressing the aforementioned limitations. Additionally, our sample is approximately 50\% larger than that of \citet{Adibekyan-21} and includes more precise measurements of planetary mass and radius. The paper is structured as follows: In Sect.\ref{sec:sample}, we introduce the sample. Sect.\ref{sec:abundances} details the methods used to determine both the primordial composition of the planet-forming disk and the composition of the exoplanets. The star-planet compositional link is discussed in Sect.\ref{sec:starplanet}, and we provide a summary of the work in  Sect.\ref{sec:summary}.

\section{The sample}                        \label{sec:sample}

\begin{figure}
\begin{center}
\includegraphics[width=1\linewidth]{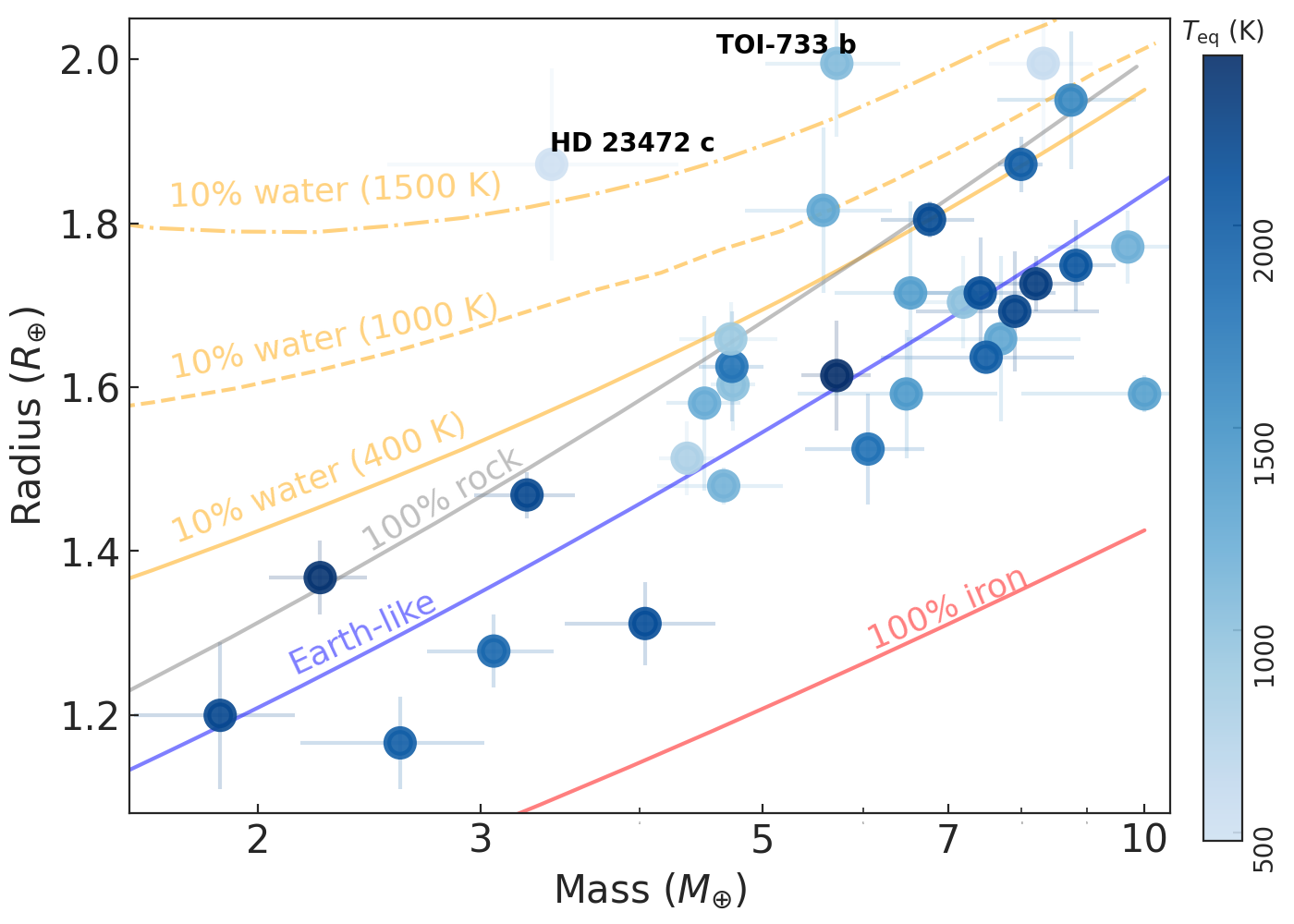}
\end{center}
\vspace{-0.4cm}
\caption{Distribution of planets in the mass--radius diagram color-coded by their $T_{\mathrm{eq}}$.  The curves display the expected mass-radius relationship for different compositions taken from \citet{Luo-24}.}
\label{r_m_plot}
\end{figure}

We started our sample selection using the PlanetS Catalog \citep{Otegi-20}\footnote{\url{https://dace.unige.ch/}}, which provides a list of transiting exoplanets with relative uncertainties in mass and radius better than 25\% and 8\%, respectively. Given our focus on potentially rocky planets, we selected only those with radii below 2 $R_{\mathrm{\oplus}}$ to exclude likely gas-rich sub-Neptunes \citep{Fulton-17, Owen-17, Venturini-20}, orbiting F, G, and K-type stars with effective temperatures ($T_{\mathrm{eff}}$) ranging between 4600 and 6500 K.  This temperature range allows for the precise determination of stellar atmospheric abundances of relevant elements. We excluded Kepler-36\,b and Kepler-36\,c which have masses determinations via transit timing variation (TTV) method as there might be a discrepancy in the density distributions of planets with mass determinations via TTV and Radial Velocity (RV) method \citep[e.g.][]{Mills-17, Leleu-23, Adibekyan-24}. Additionally, we excluded planets orbiting Kepler-65, TOI-1807, TOI-1416, and TOI-1444 due to the unavailability of high-resolution spectra for their host stars. This resulted in a sample of 32 exoplanets orbiting 30 stars, for which we could obtain stellar abundances determined from high-resolution spectra. The distribution of these planets on the Mass-Radius diagram is shown in Figure~\ref{r_m_plot}, color-coded by their equilibrium temperatures ($T_{\mathrm{eq}}$), as reported in the PlanetS Catalog. The properties of the planets are presented in Table\,\ref{tab:planet_properties}.

\section{Present-day and primordial compositions}     \label{sec:abundances}

When conducting stellar spectroscopy, the present-day properties of stars are determined. However, certain properties, including the atmospheric chemical abundances, can change as stars age, even while they remain in the main sequence. In this section, we describe how we determined the primordial compositions of stars hosting planets, and how we used this information to compute the composition of the building blocks of those planets.

\subsection{Present-day abundances of the host stars}            \label{subsec:host_composition}

Given the importance of a consistent characterization of the host stars, we compiled the stellar parameters and chemical abundances of the sample stars from five published articles where these properties were derived using the same methodology and specific tools, making them directly comparable and usable as a single set. This methodology was also applied to K2-131, EPIC\,249893012, and HD\,20329 in this study. We re-derived the parameters and abundances of EPIC\,249893012 and HD\,20329 using higher quality spectra obtained with the ESPRESSO spectrograph \citep{Pepe-21}. These parameters and their corresponding references are listed in Table~\ref{table:params_abundances}.

In all cases, the stellar atmospheric parameters ($T_{\mathrm{eff}}$, surface gravity $\log g$, microturbulent velocity $V_{\mathrm{tur}}$, and metallicity [Fe/H]) and the abundances were determined following the methods outlined in our previous works \citep[e.g.][]{Santos-13, Sousa-14, Adibekyan-12, Adibekyan-15}. In our local thermodynamic equilibrium (LTE) analysis, we measured the equivalent widths (EW) of iron lines using the ARES v2 code\footnote{The latest version of the ARES code (ARES v2) can be downloaded at \url{https://github.com/sousasag/ARES}} \citep{Sousa-15}, ensuring excitation and ionization equilibrium. We employed a grid of Kurucz model atmospheres \citep{Kurucz-93} and the 2014 version of the radiative transfer code MOOG \citep{Sneden-73}. The abundances of Mg and Si were derived using the classical curve-of-growth analysis method under the LTE assumption. Although the EWs of the spectral lines were automatically measured with ARES, we visually inspected the EW measurements to ensure accuracy.

Since determining the abundances of C and O is particularly challenging for stars cooler than about 5200 K, we decided to maintain homogeneity by predicting the abundances of these elements using a machine-learning model developed in \citet{Adibekyan-24}. This model predicts the abundances of C and O based on the input abundances of Fe, Si, and Mg wth a precision of about 0.07 dex.

\subsection{Primordial abundances of the host stars}            \label{subsec:host_primordial_composition}

We used a grid-based modeling approach (similar to \citealt{deal21}) to determine the ages and primordial abundances of the planet hosting stars. The grid of stellar models is computed with the public Code d'Evolution Stellaire Adaptatif et Modulaire\footnote{\url{https://www.ias.u-psud.fr/cesam2k20/home.html}} \citep[Cesam2k20,][]{morel08,marques13,deal18}. The parameter space of the grid varies in mass, primordial iron abundance [Fe/H]$_{prim}$ and helium mass fraction $Y_{prim}$, and the core overshoot parameter $\alpha_\mathrm{ov}$. In terms of chemical transport, the grid includes atomic diffusion with radiative accelerations \citep{deal18}, the transport induced by rotation \citep{marques13}, and we added an extra turbulent mixing calibrated on the Sun to reproduce its lithium surface abundance. The transport of chemicals is followed for H, He, Li, Be, B, C, N, O, Ne, Na, Mg, Si, Al, S, Ca, Fe, and some of their isotopes. The input physics is the same as the models presented in \citet[][]{deal20} for atomic diffusion, rotation and the microphysics, except that the primordial metal mixture follows \cite{asplund21}, the $T(\tau)$ relation in the atmosphere follows \cite{vernazza81}, and the convection formalism follows \cite{bohm58} with a solar-calibrated mixing-length parameter $\alpha_\mathrm{MLT}=1.998$. 

The stellar property inferences are performed with the Bayesian Asteroseismic Inference on a Massive Scale public code\footnote{\url{https://lesia.obspm.fr/perso/daniel-reese/spaceinn/aims/}} \citep[AIMS,][]{rendle19}. The stellar properties are inferred using $\log g$, $T_\mathrm{eff}$ and [Fe/H] as constraints, in a configuration of the code where no seismic constraints are required. Note that the observed [Fe/H] is directly compared to the one predicted by the models and not to the total metallicity [M/H] for a better accuracy \citep{deal18,moedas22}. From the posterior distributions obtained for each star, we derive the fundamental stellar properties (mass, radius, and age) and the primordial chemical composition from each elements followed by the code. The provided error bars are the 16$^\mathrm{th}$ and 84$^\mathrm{th}$ percentiles of the distributions.

The ages and primordial abundances of the sample stars are presented in Table\,\ref{tab:init_abundances}.

\begin{figure}
\begin{center}
\includegraphics[width=1\linewidth]{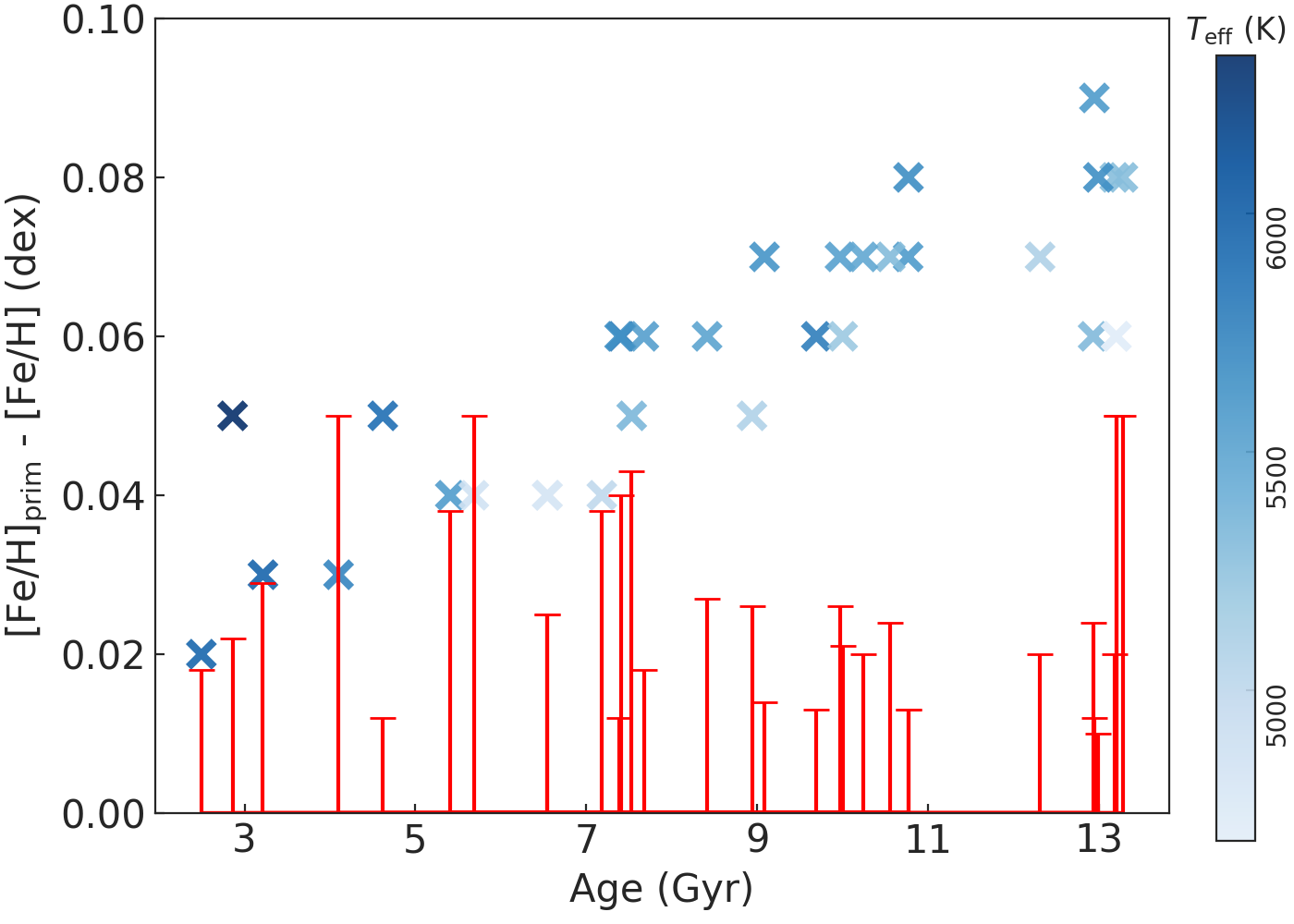}
\end{center}
\vspace{-0.4cm}
\caption{Difference (crosses) between primordial and present-day abundances of iron as a function of age and color-coded by $T_{\mathrm{eff}}$. The red errorbars represent the uncertainties of present-day [Fe/H].}
\label{fe_difussion_colorcoded_plots}
\end{figure}

In Fig.~\ref{fe_difussion_colorcoded_plots}, we display the amplitude of the iron abundance change as a function of the age and temperature of the stars. The figure illustrates that the oldest stars experience the most significant reduction in iron abundance and that, for a given age, atomic diffusion is more pronounced in hotter stars due to their narrower convective envelopes. Furthermore, the figure highlights that the temporal change in metallicity can be several times larger than the uncertainties in the present-day [Fe/H] measurements.

\begin{figure}
\begin{center}
\includegraphics[width=0.8\linewidth]{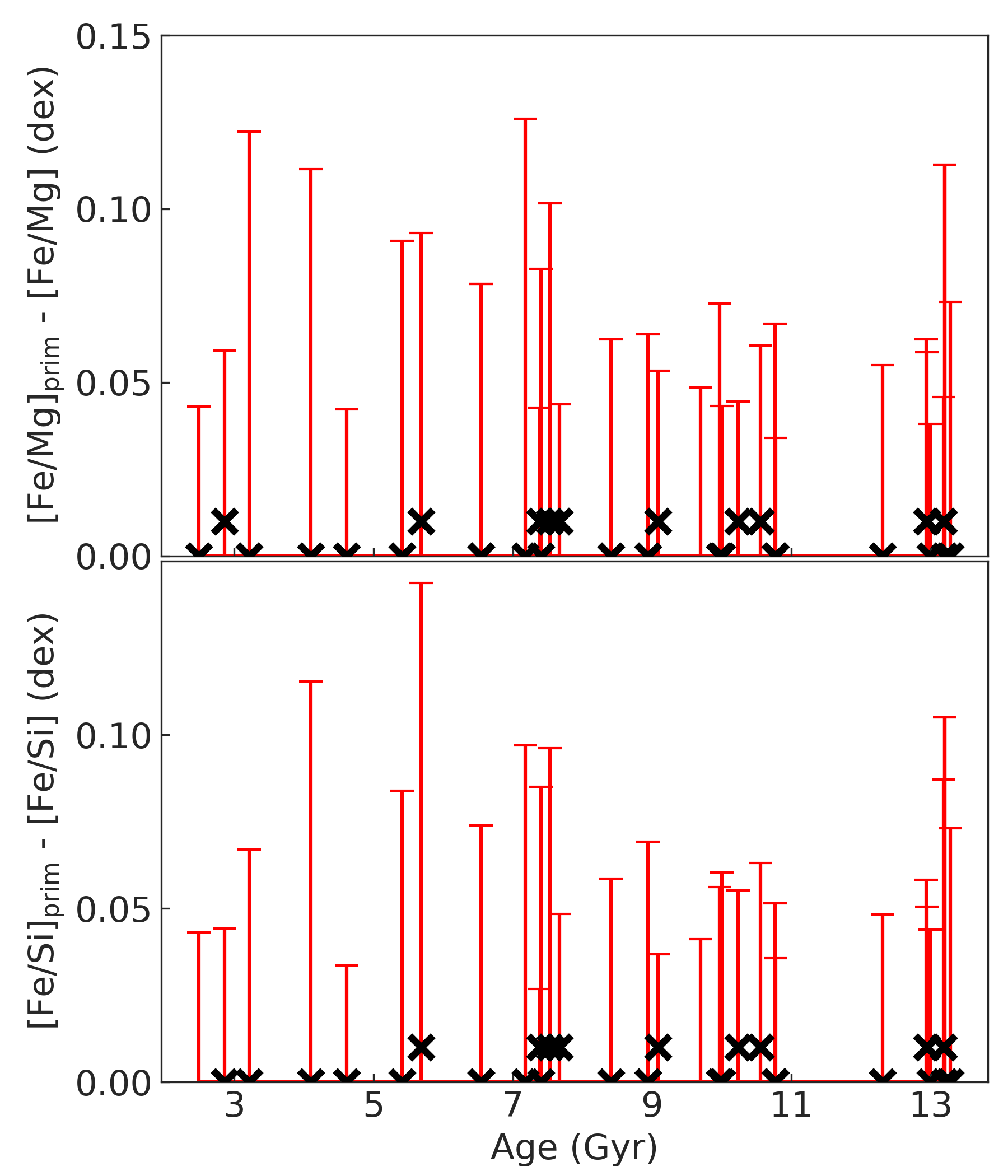}
\end{center}
\vspace{-0.4cm}
\caption{Difference (crosses) between primordial and present-day abundance ratios of [Fe/Si] and [Fe/Mg] as a function of age. The red errorbars represent the uncertainties of present-day [Fe/Si] and [Fe/Mg].}
\label{femgsi_difussion_plots}
\end{figure}

The overall trend depicted in Fig.\ref{fe_difussion_colorcoded_plots} is consistent when considering other elements used in this work. This consistency is important as it suggests that the changes in abundance ratios remain negligible, as illustrated in Fig.\ref{femgsi_difussion_plots}.

It is important to note that the impact of atomic diffusion generally depends on the competing transport processes included in the models \citep{moedas22}. A different treatment of rotation or additional mixing could also influence the results \citep[see e.g.][]{deal20}. Furthermore, for hotter stars above $\sim$6500 K, atomic diffusion may cause more pronounced changes in abundance ratios. However, since we have added extra mixing calibrated to reproduce the Sun’s lithium abundance, any variations in the efficiency of rotation or diffusion would be compensated by this calibration, resulting in a negligible impact on our results.

\subsection{Composition of planet building blocks}    \label{subsec:planet_buil_dblock_compos}

We converted the derived primordial atmospheric abundances of C, O, Mg, Si, and Fe into absolute abundances \citep{Adibekyan-19} using solar reference values from \citet{asplund21} and assuming a hydrogen atom number $\epsilon_{\mathrm{H}}$ of 10$^{12}$. These absolute stellar abundances were then used to estimate the iron-to-silicate mass fraction ($f_{\mathrm{iron}}^{\mathrm{star}}$) and the water mass fraction ($f_{\mathrm{water}}^{\mathrm{star}}$) of the protoplanetary disks using the stoichiometric model described in \citet{Santos-15, Santos-17}. Note that $f_{\mathrm{iron}}^{\mathrm{star}}$ does not depend on water availability. For the protosolar disk, the model predicts an $f_{\mathrm{iron}}^{\mathrm{\oplus}}$ of 32.3$\pm$2.1\% and an $f_{\mathrm{water}}^{\mathrm{\oplus}}$ of 58.2$\pm$3.2\% using solar abundances from \citet{asplund21}.

The present-day and primordial compositions of planet building blocks are displayed in Table\,\ref{tab:building_blocks}.

\begin{figure}
\begin{center}
\includegraphics[width=0.9\linewidth]{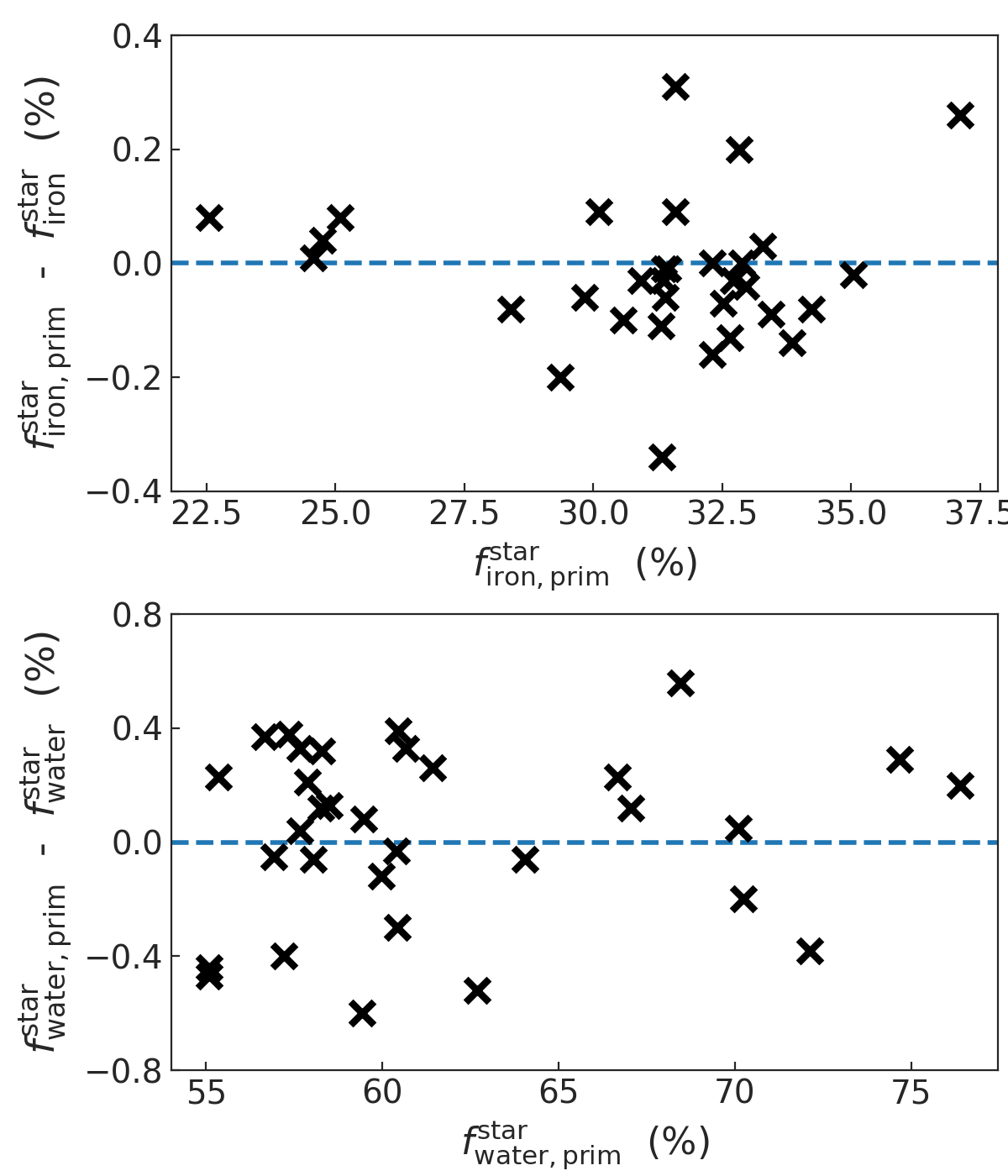}
\end{center}
\vspace{-0.4cm}
\caption{Difference in iron-to-silicate mass (top) and water mass (bottom) fractions of planet building blocks determined from the present-day and primordial stellar abundances. The differences are plotted as a function of $f_{\mathrm{iron,prim}}^{\mathrm{star}}$ and $f_{\mathrm{water,prim}}^{\mathrm{star}}$, respectively.}
\label{firon_water_difussion_plots}
\end{figure}

In Fig.\ref{firon_water_difussion_plots}, we illustrate the impact of using primordial stellar abundances instead of present-day values on $f_{\mathrm{iron}}^{\mathrm{star}}$ and $f_{\mathrm{water}}^{\mathrm{star}}$. The largest difference in the iron-to-silicate mass fraction is approximately 0.4\%, with the lowest uncertainty for this parameter around 1\% and a mean uncertainty of about 2.2\%. A similar trend is observed for the water mass fraction, with the largest difference of about 0.8\%, which is approximately 6 times and 8 times smaller than the minimum and mean uncertainty of this parameter, respectively. This behavior is consistent with the results from Sect.~\ref{subsec:host_primordial_composition}, where we demonstrated that while individual element abundances experience significant temporal changes, their ratios remain relatively stable over time. Consequently, when using Fe/Mg and Fe/Si mineralogical ratios to determine the composition of planet-building blocks, it is generally acceptable to use present-day stellar atmospheric abundances for these low temperature stars.

\begin{figure}
\begin{center}
\includegraphics[width=0.85\linewidth]{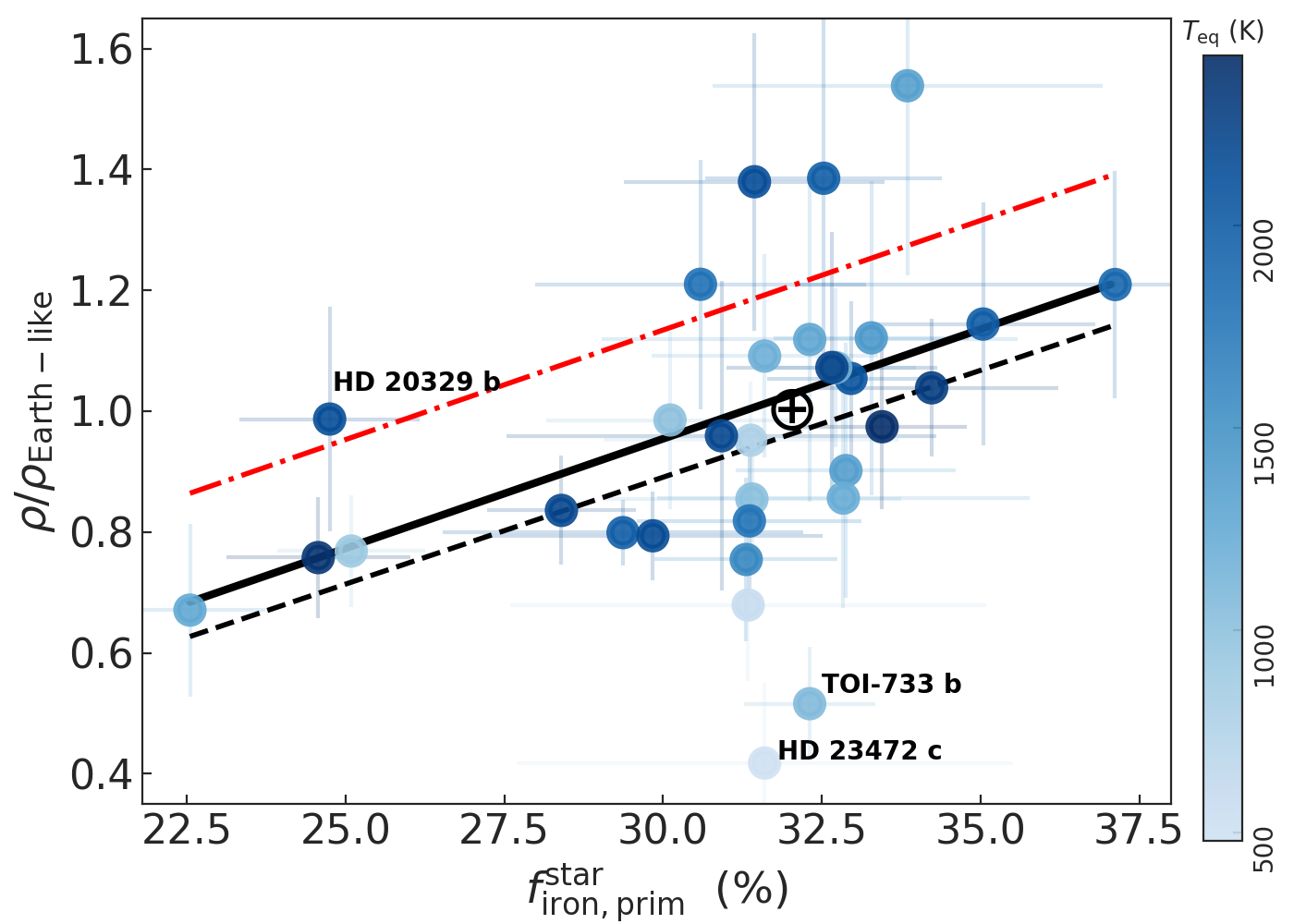}\\
\includegraphics[width=0.85\linewidth]{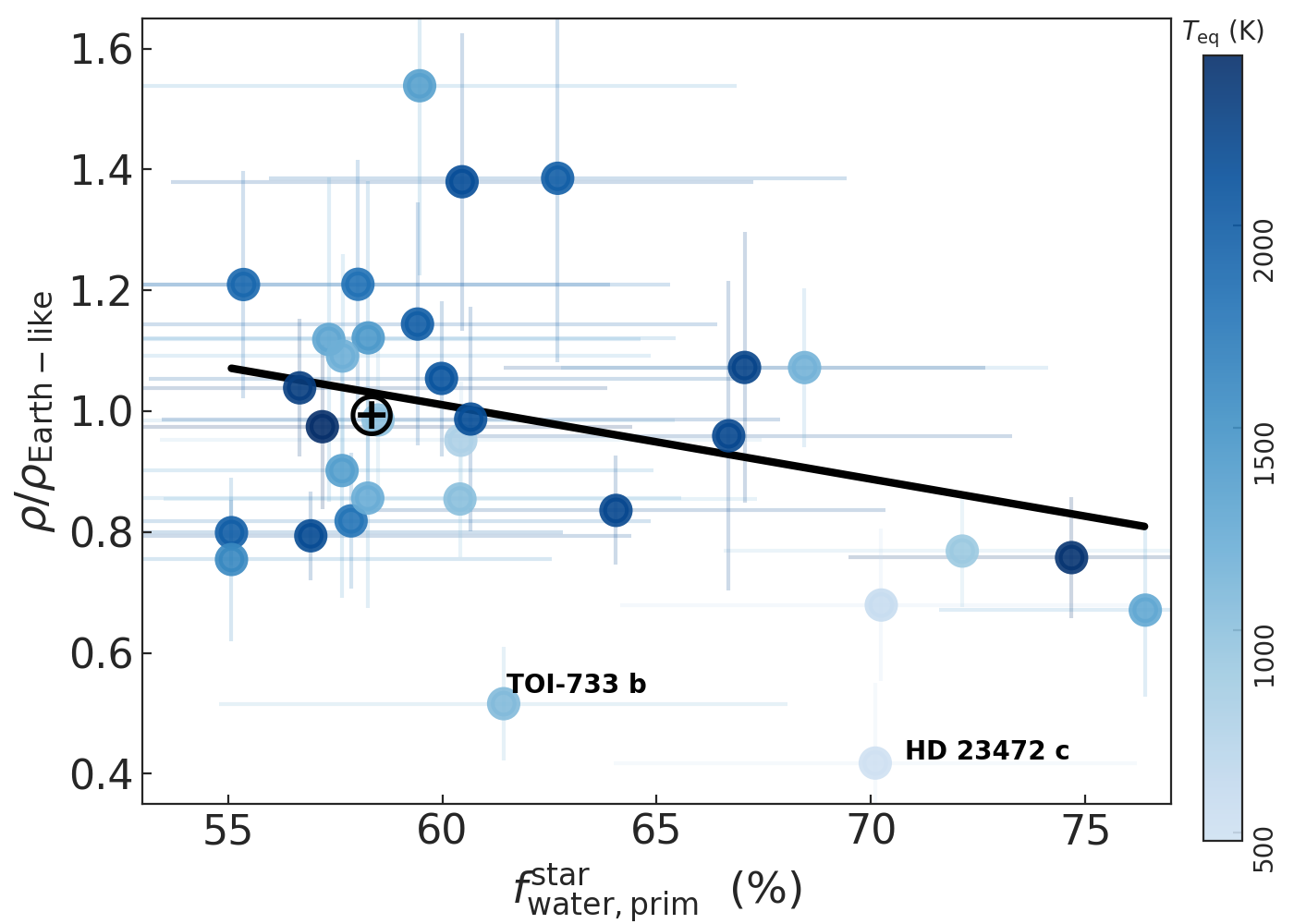}\\
\end{center}
\vspace{-0.4cm}
\caption{Scaled density of planets as a function of $f_{\mathrm{iron,prim}}^{\mathrm{star}}$ (top) and $f_{\mathrm{water,prim}}^{\mathrm{star}}$ (bottom) color-coded by the equilibrium temperature of the planets. The positions of Earth is indicated with its symbols in black. The black solid lines represent the results of OLS regression without considering TOI-733\,b and HD\,23472\,c. The red dashed-dotted line represent the visual separation between super-Earths and super-Mercuries. The black dashed line represent the result of the linear regression without considering TOI-733\,b, HD\,23472\,c and the super-Mercuries located above the red line (p-value of about 10$^{-5}$).}
\label{density_firon_fwater_colorcoded_plots}
\end{figure}

In Fig.~\ref{density_firon_fwater_colorcoded_plots}, we show how the scaled densities of planets ($\rho$/$\rho_{\mathrm{Earth-like}}$) depend on the primordial iron-to-silicate mass fraction ($f_{\mathrm{iron,prim}}^{\mathrm{star}}$) and the primordial water mass fraction ($f_{\mathrm{water,prim}}^{\mathrm{star}}$). The scaled density is calculated by dividing the bulk density ($\rho$) of the planet by the density of a planet with Earth-like composition  \citep[$\rho_{\mathrm{Earth-like}}$, ][]{Luo-24} for the same planetary mass. Throughout this paper, unless stated otherwise, we performed Ordinary Least Squares (OLS) regression using the \texttt{statsmodels} package to quantify relationships. The slopes and associated uncertainties were then used in t-statistics (via \texttt{SCIPY}) to test the null hypothesis that the data can be modeled with a zero slope. The null hypothesis is rejected if the resulting two-tailed p-values (P(t-stat)) are below the typical threshold of 0.05. We opted for OLS over Orthogonal Distance Regression (ODR), which was used in \citet{Adibekyan-21}, based on the discussion in \citet{Brinkman-24}, which concluded that there is a non-negligible probability of obtaining a statistically significant relation when using ODR on a sample similar to ours.

Compared to \citet{Adibekyan-21}, the top panel of Fig.\,\ref{density_firon_fwater_colorcoded_plots} shows two low-density planets (TOI-733\,b and HD\,23472\,c) orbiting stars with solar $f_{\mathrm{iron,prim}}^{\mathrm{star}}$. In Fig.\,\ref{r_m_plot}, these two planets are the farthest from the Earth-like composition curve, and our interior analysis suggests that they contain a significant amount of water (above 10\%, see Table\,\ref{tab:planet_composition}), potentially in the form of steam. We excluded these two planets from the regression analysis shown in Fig.~\ref{density_firon_fwater_colorcoded_plots}. It is important to note that $f_{\mathrm{iron,prim}}^{\mathrm{star}}$ refers to the iron-to-silicate mass ratio and does not account for water in the calculation.

The top panel of Fig.\,\ref{density_firon_fwater_colorcoded_plots} depicts that low-density planets are likely formed in protoplanetary disks with a low iron-to-silicate mass fraction, with an OLS regression p-value of 0.002 suggesting a statistically significant correlation between scaled density and $f_{\mathrm{iron,prim}}^{\mathrm{star}}$. This correlation was previously noted in \citet{Adibekyan-21}. It is important to distinguish that $f_{\mathrm{iron,prim}}^{\mathrm{star}}$ and [Fe/H] of the host stars are different parameters. Furthermore, there is no statistically significant correlation (p-value $\sim$ 0.08) between scaled density and [Fe/H].

The bottom panel of Fig.~\ref{density_firon_fwater_colorcoded_plots} suggests a possible negative correlation between scaled density and $f_{\mathrm{water,prim}}^{\mathrm{star}}$. However, this trend is not statistically significant, with a p-value of 0.07. Overall, both low iron content and high water content are likely factors contributing to the low density of rocky planets.

Interestingly, the density gap between super-Mercuries and super-Earths appears less pronounced compared to the findings of \citet{Adibekyan-21}, which is in line with the results of some resent studies \citep[][]{Brinkman-24, Schulze-24}. This discrepancy can be attributed to updated mass and radius determinations and stronger precision constraints on these parameters for the sample planets. Notably, the four highest-density planets in our sample orbit stars with above-average $f_{\mathrm{iron,prim}}^{\mathrm{star}}$. 

\begin{figure}
\begin{center}
\includegraphics[width=0.85\linewidth]{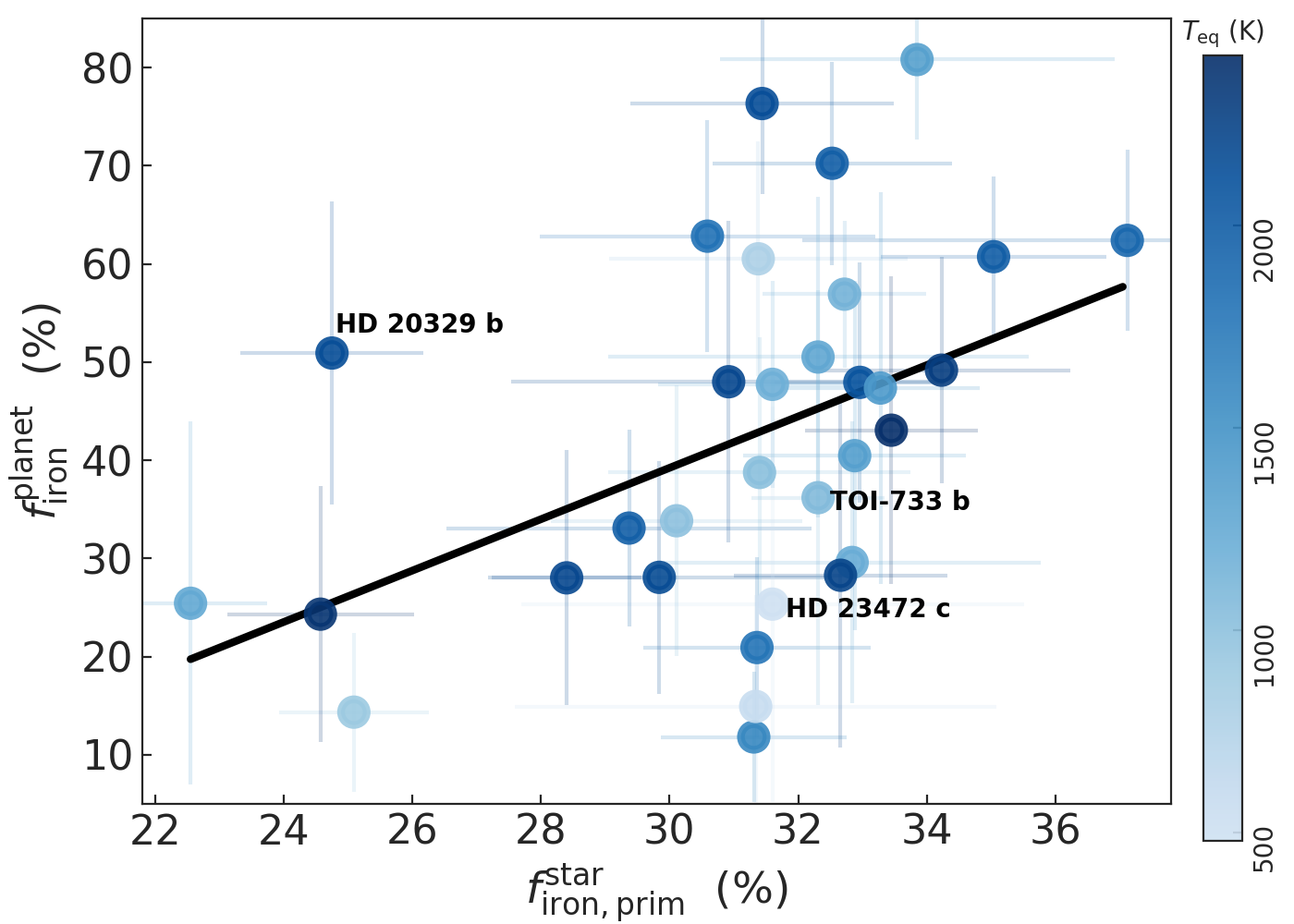}\\
\includegraphics[width=0.85\linewidth]{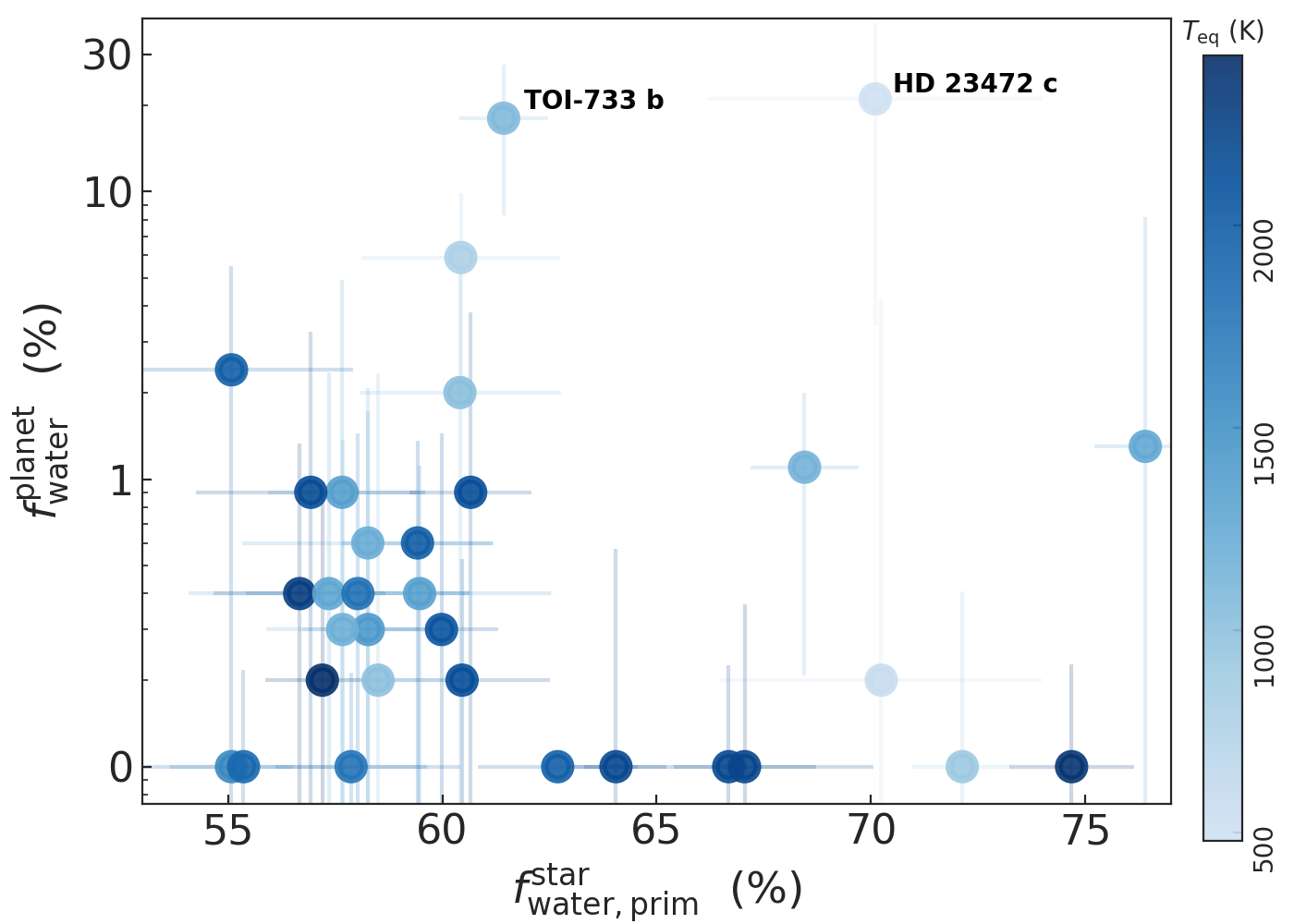}\\
\end{center}
\vspace{-0.4cm}
\caption{Iron-to-silicate and water mass fractions of planets and their host stars color-coded by the equilibrium temperature of the planets.  The black solid line represents the results of OLS regression.}
\label{firon_fwater_star_planet}
\end{figure}

Additionally, HD\,20329\,b, which orbits a star with a very low $f_{\mathrm{iron,prim}}^{\mathrm{star}}$, has a scaled density similar to that of Earth. This suggests that HD\,20329\,b might have a significantly higher $f_{\mathrm{iron}}^{\mathrm{planet}}$ (see. Fig.~\ref{firon_fwater_star_planet}) than what would be expected from its host star’s protoplanetary disk composition, potentially classifying it as a super-Mercury \citep{Schulze-21}. \citet{Adibekyan-21} previously proposed that the formation of high-density super-Mercuries might be linked to the composition of the host star, a hypothesis that has been further explored theoretically in \citet{Johansen-22, Mah-23}.

\subsection{Interior composition of exoplanets}    \label{subsec:planet_composition}

To compare stellar abundances with planetary abundances, we use inference methods to estimate the confidence regions for the planetary iron-to-silicate mass fraction ($f_{\mathrm{iron}}^{\mathrm{planet}}$), the water mass fraction ($f_{\mathrm{water,prim}}^{\mathrm{star}}$), and the iron-to-magnesium ratio ($(Fe/Mg)_{\mathrm{planet}}$).

We use an interior model with 3 components: core, mantle, and water (mainly steam).  The model is based on \citet{Dorn-15} with updates from \citet{Luo-24}. The water can be present in the core, the mantle, and the surface, depending on the specific thermal state of the planets.
In brief, the interior model uses different sets of equation of states for the iron in the core \citep{hakim_new_2018,miozzi_new_2020,Luo-24}, for silicates in the mantle \citep{fischer_equation_2011, faik_equation_2018, hemley_constraints_1992, musella_physical_2019, connolly}, and surface water \citep[mainly in steam phase for the planets of interest][]{haldemann_aqua_2020}. We assume adiabatic temperature profiles in the core, mantle and steam layer, except for pressures below 0.1 bar in the steam layer where we assume an isothermal profile. Additionally, we assumed a pure iron core and fixed the Mg/Si number ratio to 1.04, which is representative of Earth-like composition. The planet radius is set at 1 mbar.

Water can be added to the mantle melts, while the solid mantle is assumed to be dry. The addition of water reduces the density, for which we follow \citet{bajgain_structure_2015}. For small water mass fractions, this reduction is nearly independent of pressure and temperature. Further, water can be present in the core for which we use the equation of states of \citep{Luo-24}, that accounts for the reduction in density. Further details on the model can be found in \citet{Luo-24}.

The composition of the planets are presented in Table\,\ref{tab:planet_composition}.

\section{Star--planet compositional link}     \label{sec:starplanet}

Having independently determined the iron-to-silicate mass fractions from both the stellar composition, and the planetary mass and radius, we can now examine the compositional link between solar-type stars and their low-mass planets. The top panel of Fig.~\ref{firon_fwater_star_planet} shows the relationship between planetary $f_{\mathrm{iron}}^{\mathrm{planet}}$ and stellar $f_{\mathrm{iron,prim}}^{\mathrm{star}}$. The OLS regression analysis, with a slope of 2.6$\pm$0.9 and p-value of 0.009, indicates a significant correlation between these two parameters. Given the ongoing discussion in the literature about this relationship, and to ensure the robustness of this result, we applied different regression methods  (Appendix\,\ref{apdx:relation}), accounting for the uncertainties in the parameters. While the exact form of the relation varies with the methodology used, the presence of a steeper-than-one-to-one correlation remains evident. The figure also reveals that the range of planetary iron-to-silicate mass fractions is significantly larger than that of the protoplanetary disk, confirming the previous findings \citep{Plotnykov-20,Adibekyan-21}.

In the bottom panel of Fig.~\ref{firon_fwater_star_planet}, we display the relationship between planetary and protoplanetary water mass fractions. The absence of a clear correlation is somewhat expected, given that $f_{\mathrm{water,prim}}^{\mathrm{star}}$ is calculated with the assumption that the planet-building blocks are located outside the water snowline. The low water content observed in most of the studied planets may actually suggest that they formed inside the water snowline. Several recent studies have also suggested a low water content in close-in super-Earths \citep{Muller-24, Rogers-24}.

\begin{figure}
\begin{center}
\includegraphics[width=0.85\linewidth]{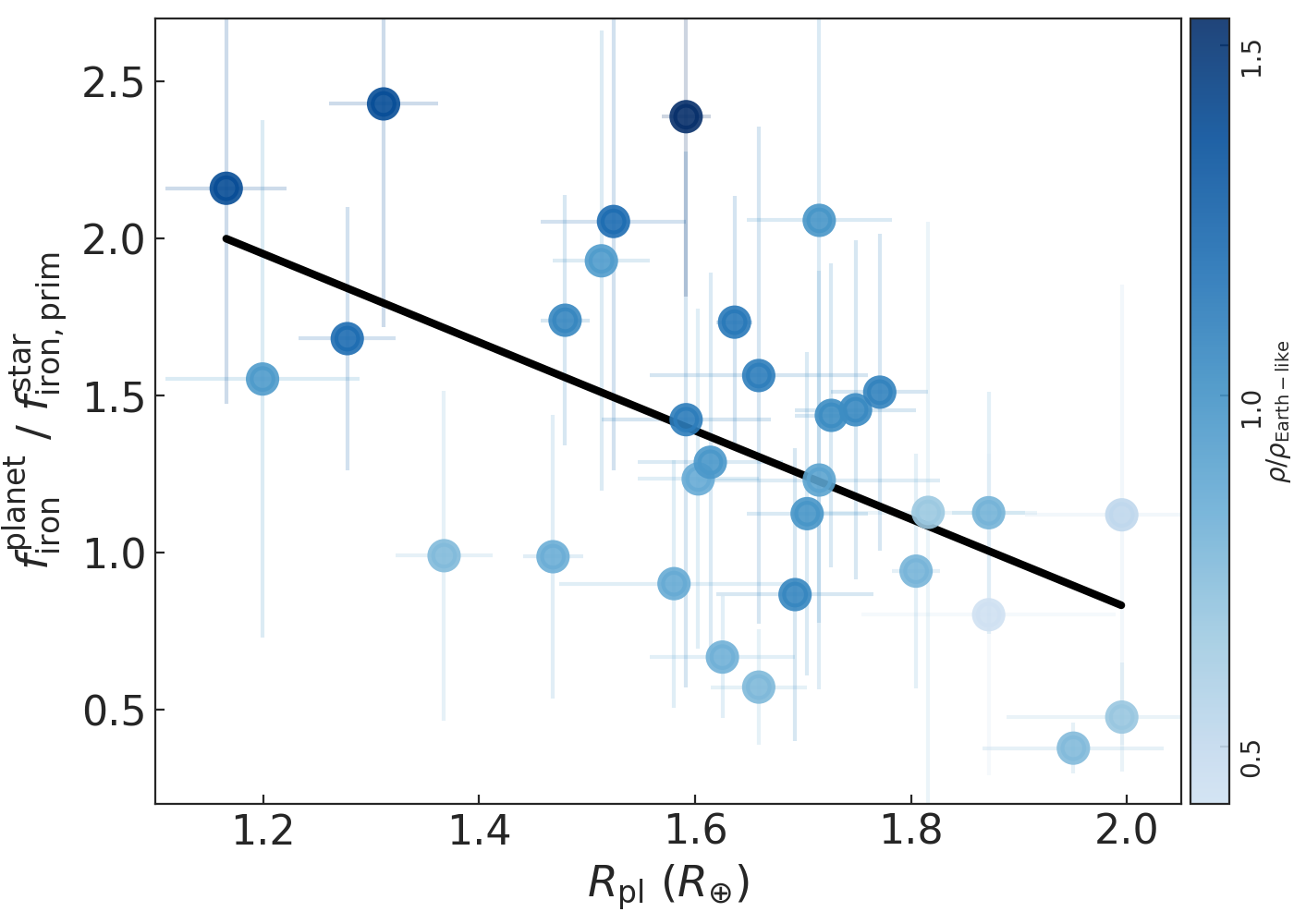}\\
\includegraphics[width=0.85\linewidth]{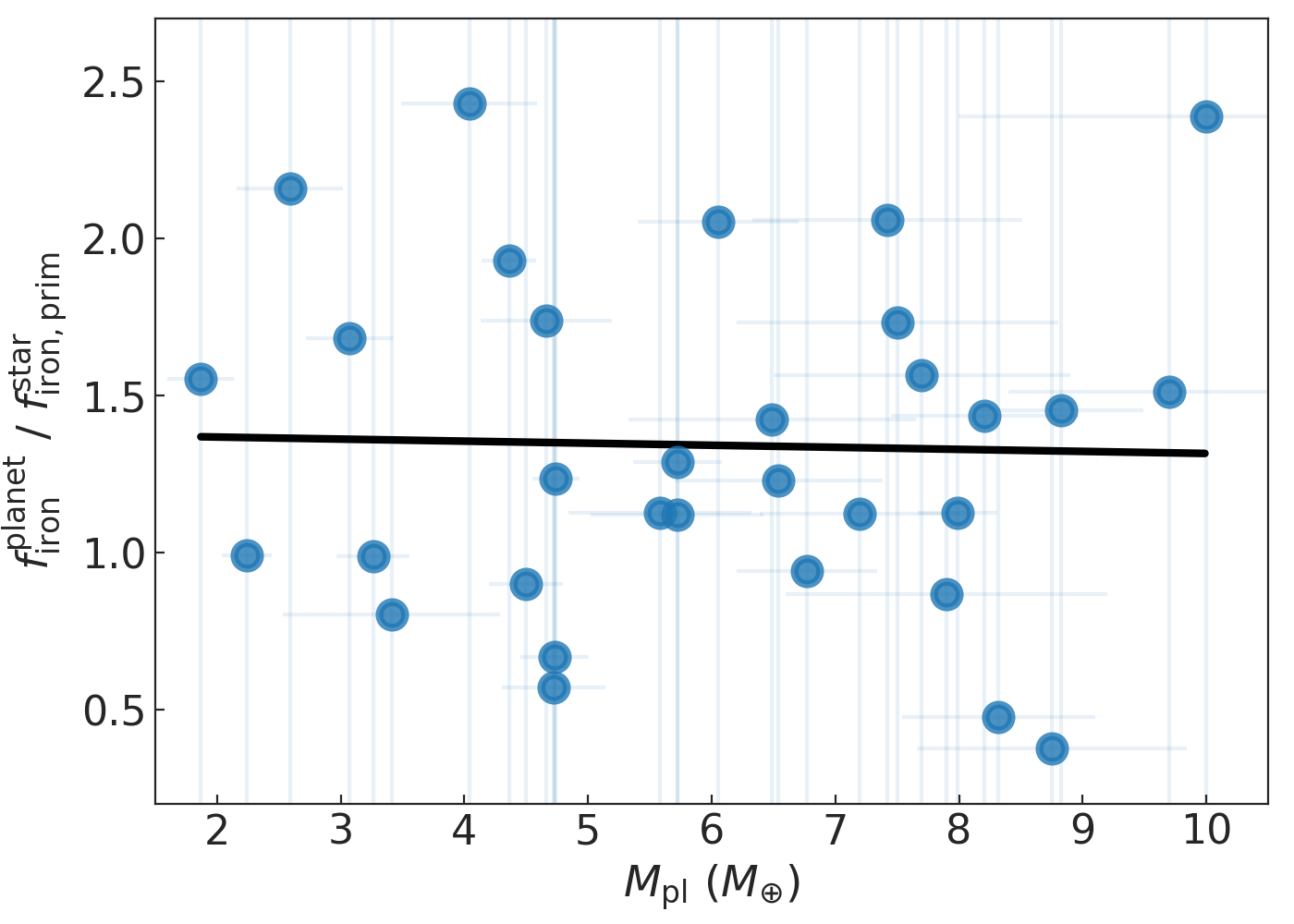}\\
\end{center}
\vspace{-0.4cm}
\caption{The ratio of iron-to-silicate mass fraction of planets and the planet building blocks as a function of planetary radius (top panel) and mass (bottom panel). The black solid lines represent the results of the linear regression. The symbols in the top panel are color-coded by the scaled density of the planets.}
\label{firon_star_planet_R_M}
\end{figure}

We further explore the planet-star compositional link by examining the ratio of $f_{\mathrm{iron}}^{\mathrm{planet}}$ to $f_{\mathrm{iron,prim}}^{\mathrm{star}}$ as a function of planetary radius (top panel of Fig.\ref{firon_star_planet_R_M}) and planetary mass (bottom panel of Fig.\ref{firon_star_planet_R_M}). There is a clear correlation between this ratio and planetary radius (p-value $<$ 0.001), and no correlation with  planetary mass (p-value of about 0.9). This is most likely due to the fact that planetary density depends on radius with a cubic relation, while the dependence on mass is only linear. However, the top panel also shows that for a given radius, denser planets have a higher ratio of $f_{\mathrm{iron}}^{\mathrm{planet}}$ to $f_{\mathrm{iron,prim}}^{\mathrm{star}}$.

\begin{figure}
\begin{center}
\includegraphics[width=0.9\linewidth]{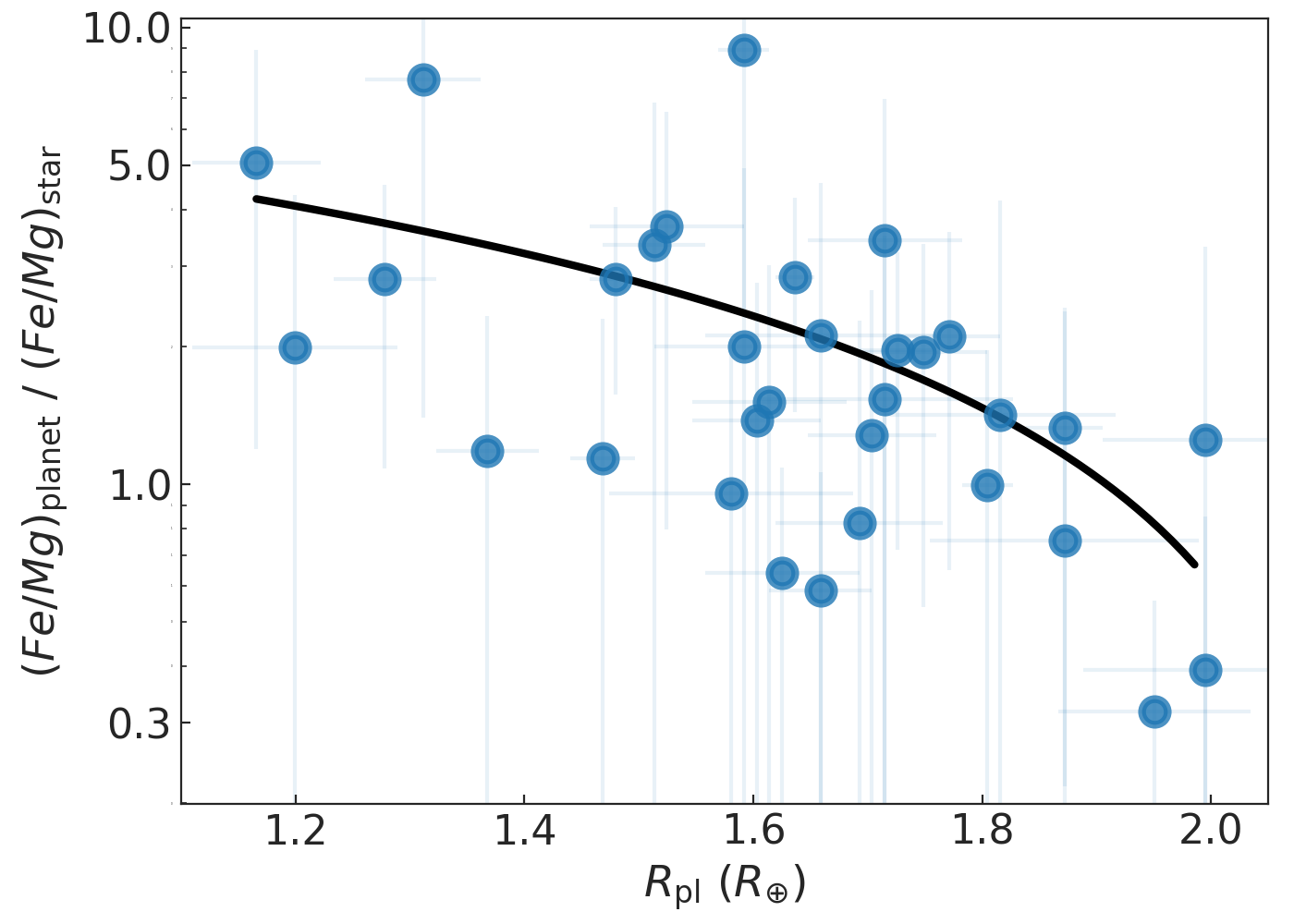}
\end{center}
\vspace{-0.4cm}
\caption{The Fe/Mg abundance ratio of planets and their host stars as a function of planetary radius. The black solid line in the plot represent the results of the OLS regression (p-value $\sim$ 0.006) performed on the linearly scaled data. }
\label{FeMg_FeO_R}
\end{figure}

As planetary interiors are often modeled under the assumption that the Fe/Mg ratio in planets is similar to that of their host stars, Fig.\ref{FeMg_FeO_R} illustrates how the Fe/Mg planet-to-host ratio varies with planetary radius. The figure reveals that the Fe/Mg ratio in planets can be up to three times lower or up to ten times higher than in their host stars, indicating a large variation between the formed planets and the protoplanetary disk. This substantial variation suggests that processes occurring within the protoplanetary disk during planet formation play a significant role in determining the final composition of rocky planets. Additionally, Fig.\ref{FeMg_FeO_R} shows a clear dependence of the $(Fe/Mg){\mathrm{planet}}$ / $(Fe/Mg){\mathrm{star}}$ ratio on planetary radius. 

\section{Summary}     \label{sec:summary}

In this work, we analyzed a sample of 32 low-mass planets orbiting 30 FGK-type stars to investigate the compositional link between stars and their planets. The selected planets from the PlanetS Catalog have relative uncertainties in mass and radius better than 25\% and 8\%, respectively. Compared to our previous study \citep{Adibekyan-21}, the current sample includes about 50\% more planets, and our analysis was enhanced by determining the primordial compositions of the hosts and using a planetary interior model that accounts also for presence water (primarily in steam form).

We derived the primordial stellar abundances of C, O, Mg, Si, and Fe from their present-day atmospheric values, demonstrating that while individual elemental abundances can vary significantly over time, the ratios between them remain relatively stable. Note that this may not be true for hotter stars, and that the exact values are subject to the assumptions and limitations of the modeling.

We then used the primordial abundances as input for the stoichiometric model presented in \citet{Santos-15} to estimate the iron-mass fraction ($f_{\mathrm{iron,prim}}^{\mathrm{star}}$) and water-mass fraction ($f_{\mathrm{water,prim}}^{\mathrm{star}}$) of the protoplanetary disks. When comparing the present-day and primordial compositions of the planet-building disks, we found negligible differences, as the abundance ratios of Fe/Mg and Fe/Si remain relatively constant over the evolution of stars in the main sequence. This consistency suggests that using present-day stellar abundances to infer the composition of protoplanetary disks is a valid approach, as the key elemental ratios critical for planet formation do not change significantly over time.

Our analysis revealed a strong correlation between the scaled density of rocky planets and the primordial iron-to-silicate mass fraction of their protoplanetary disks ($f_{\mathrm{iron,prim}}^{\mathrm{star}}$). Although the density gap between super-Mercuries and super-Earths is less pronounced than previously observed in \citet{Adibekyan-21}, it remains visible.

We employed a three-component planetary interior model \citep{Dorn-15, Luo-24} to derive the compositions of the planets based on their mass and radius. While we found a statistically significant correlation between the iron-to-silicate mass fraction of planets and their host stars, no such correlation was observed for the water mass fraction. This lack of correlation may be due to the assumption in the stoichiometric model that the building blocks of planets formed outside the water snowline, while the observed low water content suggests formation inside the snowline.

As already found in \citet{Adibekyan-21}, the range of $f_{\mathrm{iron}}^{\mathrm{planet}}$ is significantly broader than that of their building blocks. This indicates that while the primordial composition of the protoplanetary disk sets a baseline for planetary composition, various processes during planet formation and evolution can lead to substantial deviations in the iron-to-silicate mass fraction of the resulting planets compared to their original building materials.

Additionally, we compared the Fe/Mg ratios in planets and their host stars, finding a variation from 0.3 to 10 times between the two. This ratio was also found to correlate with planetary radius, further indicating that processes within the protoplanetary disk significantly influence the final composition of rocky planets.

In conclusion, we confirm the previous results \citep{Adibekyan-21} suggesting that while a correlation exists between the composition of rocky planets and their host stars, the assumption of a strict one-to-one elemental abundance relationship is not supported by our observations.

\begin{acknowledgements}
This work was supported by Funda\c{c}\~ao para a Ci\^encia e Tecnologia through national funds and by FEDER through COMPETE2020 - Programa Operacional Competitividade e Internacionalização by these grants: UIDB/04434/2020; UIDP/04434/2020; 2022.06962.PTDC. Funded/Co-funded by the European Union (ERC, FIERCE, 101052347). Views and opinions expressed are however those of the author(s) only and do not necessarily reflect those of the European Union or the European Research Council. Neither the European Union nor the granting authority can be held responsible for them. TLC is supported by Funda\c c\~ao para a Ci\^encia e a Tecnologia (FCT) in the form of a work contract (CEECIND/00476/2018).

We would like to thank the referee, Joe Schulze, for the constructive comments and suggestions, which have significantly improved both the quality and clarity of our work.

This research has made use of the NASA Exoplanet Archive operated by the California Institute of Technology under contract with the National Aeronautics and Space Administration under the Exoplanet Exploration Program.

In this work we used the Python language and several scientific packages: Numpy \citep{van_der_Walt-11}, Scipy \citep{Virtanen-20}, Pandas \citep{mckinney-proc-scipy-2010}, statsmodels \citep{seabold-proc-scipy-2010}, and Matplotlib \citep{Hunter-07}.
\end{acknowledgements}

\bibliography{references.bib}

\clearpage

\begin{appendix}

\section{Tables with compositions of stars and their planets}

\begin{table*}[]
    \caption{\label{tab:planet_properties} The main properties of the planets. }
    \centering
    \small
    \begin{tabular}{lllll}
    \hline\hline
        Planet & M ($M_{\mathrm{\oplus}}$) & R ($R_{\mathrm{\oplus}}$) & $\rho$/$\rho_{\mathrm{Earth-like}}$ & $T_{\mathrm{eq}}$ (K) \\ \hline
        55 Cnc e & 7.989$_{-0.33}^{+0.32}$ & 1.87$_{-0.034}^{+0.034}$ & 0.80$\pm$0.05 & 1986 \\
        EPIC 249893012 b & 8.749$_{-1.08}^{+1.09}$ & 1.95$_{-0.078}^{+0.090}$ & 0.75$\pm$0.14 & 1596 \\ 
        HD 137496 b & 4.039$_{-0.55}^{+0.55}$ & 1.31$_{-0.045}^{+0.056}$ & 1.38$\pm$0.25 & 2141 \\ 
        HD 213885 b & 8.828$_{-0.65}^{+0.66}$ & 1.75$_{-0.056}^{+0.056}$ & 1.05$\pm$0.13 & 2083 \\ 
        HD 219134 b & 4.738$_{-0.19}^{+0.19}$ & 1.60$_{-0.056}^{+0.056}$ & 0.85$\pm$0.10 & 1014 \\ 
        HD 219134 c & 4.360$_{-0.22}^{+0.22}$ & 1.51$_{-0.045}^{+0.045}$ & 0.95$\pm$0.10 & 782 \\ 
        HD 3167 b & 4.729$_{-0.29}^{+0.28}$ & 1.63$_{-0.056}^{+0.078}$ & 0.82$\pm$0.11 & 1775 \\ 
        HD 80653 b & 5.720$_{-0.35}^{+0.36}$ & 1.61$_{-0.067}^{+0.067}$ & 0.97$\pm$0.14 & 2421 \\ 
        EPIC 220674823 b & 8.209$_{-0.74}^{+0.76}$ & 1.73$_{-0.034}^{+0.034}$ & 1.04$\pm$0.11 & 2290 \\ 
        K2-111 b & 5.581$_{-0.73}^{+0.74}$ & 1.82$_{-0.090}^{+0.112}$ & 0.67$\pm$0.14 & 1266 \\ 
        K2-229 b & 2.590$_{-0.43}^{+0.43}$ & 1.17$_{-0.045}^{+0.067}$ & 1.39$\pm$0.30 & 1980 \\ 
        K2-265 b & 6.540$_{-0.84}^{+0.84}$ & 1.71$_{-0.112}^{+0.112}$ & 0.90$\pm$0.21 & 1362 \\ 
        K2-291 b & 6.489$_{-1.16}^{+1.16}$ & 1.59$_{-0.067}^{+0.090}$ & 1.12$\pm$0.26 & 1420 \\ 
        K2-38 b & 7.700$_{-1.10}^{+1.20}$ & 1.66$_{-0.101}^{+0.101}$ & 1.12$\pm$0.27 & 1280 \\ 
        Kepler-10 b & 3.261$_{-0.30}^{+0.30}$ & 1.47$_{-0.022}^{+0.034}$ & 0.84$\pm$0.09 & 2196 \\ 
        Kepler-107 c & 9.998$_{-2.00}^{+2.00}$ & 1.59$_{-0.022}^{+0.022}$ & 1.54$\pm$0.31 & 1352 \\ 
        Kepler-20 b & 9.699$_{-1.30}^{+1.30}$ & 1.77$_{-0.034}^{+0.056}$ & 1.09$\pm$0.17 & 1183 \\ 
        Kepler-78 b & 1.869$_{-0.26}^{+0.27}$ & 1.20$_{-0.090}^{+0.090}$ & 0.96$\pm$0.26 & 2200 \\ 
        Kepler-93 b & 4.659$_{-0.53}^{+0.53}$ & 1.48$_{-0.022}^{+0.022}$ & 1.07$\pm$0.13 & 1149 \\ 
        HD 15337 b & 7.198$_{-0.81}^{+0.81}$ & 1.70$_{-0.056}^{+0.056}$ & 0.99$\pm$0.15 & 1000 \\ 
        TOI-561 b & 2.240$_{-0.20}^{+0.20}$ & 1.37$_{-0.045}^{+0.045}$ & 0.76$\pm$0.10 & 2373 \\ 
        WASP-47 e & 6.769$_{-0.57}^{+0.57}$ & 1.80$_{-0.022}^{+0.022}$ & 0.79$\pm$0.07 & 2166 \\ 
        HD 20329 b & 7.421$_{-1.09}^{+1.09}$ & 1.71$_{-0.067}^{+0.067}$ & 0.99$\pm$0.19 & 2149 \\ 
        Kepler-21 b & 7.500$_{-1.30}^{+1.30}$ & 1.64$_{-0.011}^{+0.022}$ & 1.14$\pm$0.20 & 1995 \\ 
        CoRoT-7 b & 6.054$_{-0.65}^{+0.65}$ & 1.52$_{-0.067}^{+0.067}$ & 1.21$\pm$0.21 & 1800 \\ 
        HIP 29442 c & 4.500$_{-0.30}^{+0.30}$ & 1.58$_{-0.112}^{+0.101}$ & 0.86$\pm$0.18 & 1214 \\ 
        TOI-733 b & 5.720$_{-0.68}^{+0.70}$ & 2.00$_{-0.090}^{+0.090}$ & 0.52$\pm$0.09 & 1066 \\ 
        TOI-431 b & 3.070$_{-0.35}^{+0.35}$ & 1.28$_{-0.045}^{+0.045}$ & 1.21$\pm$0.19 & 1898 \\ 
        K2-131 b & 7.901$_{-1.30}^{+1.30}$ & 1.69$_{-0.056}^{+0.090}$ & 1.07$\pm$0.22 & 2227 \\ 
        HD 23472 b & 8.320$_{-0.79}^{+0.78}$ & 2.00$_{-0.101}^{+0.112}$ & 0.68$\pm$0.13 & 570 \\ 
        HD 23472 c & 3.410$_{-0.81}^{+0.88}$ & 1.87$_{-0.112}^{+0.123}$ & 0.42$\pm$0.13 & 479 \\ 
        HD 136352 b & 4.719$_{-0.42}^{+0.42}$ & 1.66$_{-0.045}^{+0.045}$ & 0.77$\pm$0.09 & 908 \\ \hline
    \end{tabular}
\end{table*}

\begin{table*}[]
\caption{\label{tab:stellar_parameters} The stellar atmospheric parameters and present-day abundances of the planet host stars.}
\centering
\small
\begin{tabular}{llllllllll}
\hline\hline
        Star & $T_{\mathrm{eff}}$ (K) & $\log g$ (dex) & $V_{\mathrm{tur}}$ (km/s) & [Fe/H] (dex) & [Mg/H] (dex) & [Si/H] (dex) & [C/H] (dex) & [O/H] (dex) & Ref. \\ \hline
        55 Cnc & 5341$\pm$62 & 4.26$\pm$0.14 & 0.94$\pm$0.09 & 0.32$\pm$0.04 & 0.42$\pm$0.06 & 0.35$\pm$0.05 & 0.28$\pm$0.10 & 0.33$\pm$0.10 & 1 \\ 
        CoRoT-7 & 5288$\pm$27 & 4.40$\pm$0.07 & 0.90$\pm$0.05 & 0.02$\pm$0.02 & 0.03$\pm$0.03 & 0.07$\pm$0.08 & -0.01$\pm$0.10 & 0.05$\pm$0.10 & 2 \\ 
        EPIC 220674823 & 5522$\pm$34 & 4.34$\pm$0.05 & 0.87$\pm$0.05 & 0.10$\pm$0.03 & 0.07$\pm$0.05 & 0.05$\pm$0.03 & 0.00$\pm$0.10 & 0.05$\pm$0.10 & 2 \\ 
        EPIC 249893012 & 5571$\pm$17 & 4.05$\pm$0.03 & 1.03$\pm$0.03 & 0.19$\pm$0.02 & 0.23$\pm$0.05 & 0.20$\pm$0.03 & 0.11$\pm$0.10 & 0.17$\pm$0.10 & 5 \\ 
        HD 136352 & 5664$\pm$14 & 4.39$\pm$0.02 & 0.85$\pm$0.02 & -0.34$\pm$0.01 & -0.14$\pm$0.03 & -0.23$\pm$0.04 & -0.23$\pm$0.10 & 0.01$\pm$0.10  & 2 \\ 
        HD 137496 & 5799$\pm$61 & 4.05$\pm$0.10 & 1.16$\pm$0.02 & -0.03$\pm$0.04 & 0.01$\pm$0.02 & -0.03$\pm$0.03 & -0.06$\pm$0.10 & 0.02$\pm$0.10  & 2 \\ 
        HD 15337 & 5081$\pm$43 & 4.21$\pm$0.13 & 0.71$\pm$0.10 & 0.03$\pm$0.03 & 0.09$\pm$0.04 & 0.06$\pm$0.05 & 0.01$\pm$0.10 & 0.07$\pm$0.10  & 2 \\ 
        HD 20329 & 5573$\pm$16 & 4.29$\pm$0.03 & 0.85$\pm$0.03 & -0.11$\pm$0.02 & 0.08$\pm$0.05 & 0.03$\pm$0.04 & 0.01$\pm$0.10 & 0.05$\pm$0.10  & 5 \\ 
        HD 213885 & 5885$\pm$15 & 4.39$\pm$0.03 & 1.06$\pm$0.02 & 0.00$\pm$0.01 & -0.01$\pm$0.04 & -0.01$\pm$0.03 & -0.06$\pm$0.10 & 0.02$\pm$0.10 & 1 \\ 
        HD 219134 & 4789$\pm$54 & 4.15$\pm$0.19 & 0.69$\pm$0.14 & -0.03$\pm$0.03 & -0.03$\pm$0.06 & 0.00$\pm$0.06 & -0.09$\pm$0.10 & 0.02$\pm$0.10 & 1 \\ 
        HD 23472 & 4684$\pm$99 & 4.53$\pm$0.08 & 0.25$\pm$0.49 & -0.20$\pm$0.05 & -0.19$\pm$0.09 & -0.18$\pm$0.08 & -0.25$\pm$0.10 & 0.01$\pm$0.10 & 3 \\ 
        HD 3167 & 5306$\pm$36 & 4.35$\pm$0.07 & 0.69$\pm$0.06 & 0.04$\pm$0.02 & 0.10$\pm$0.04 & 0.03$\pm$0.04 & -0.01$\pm$0.10 & 0.05$\pm$0.10 & 1 \\ 
        HD 80653 & 5942$\pm$23 & 4.34$\pm$0.05 & 1.14$\pm$0.03 & 0.34$\pm$0.02 & 0.31$\pm$0.03 & 0.33$\pm$0.03 & 0.26$\pm$0.10 & 0.32$\pm$0.10 & 2 \\ 
        HIP 29442 & 5289$\pm$35 & 4.39$\pm$0.03 & 0.74$\pm$0.06 & 0.24$\pm$0.05 & 0.26$\pm$0.04 & 0.21$\pm$0.04 & 0.17$\pm$0.10 & 0.24$\pm$0.10 & 4 \\ 
        K2-111 & 5775$\pm$60 & 4.25$\pm$0.15 & 1.02$\pm$0.50 & -0.47$\pm$0.01 & -0.20$\pm$0.04 & -0.30$\pm$0.03 & -0.31$\pm$0.10 & 0.01$\pm$0.10  & 2 \\ 
        K2-131 & 5090$\pm$30 & 4.39$\pm$0.07 & 0.99$\pm$0.06 & -0.12$\pm$0.02 & -0.14$\pm$0.04 & -0.12$\pm$0.03 & -0.16$\pm$0.10 & 0.01$\pm$0.10 & 5 \\ 
        K2-229 & 5196$\pm$35 & 4.39$\pm$0.07 & 0.91$\pm$0.06 & -0.06$\pm$0.02 & -0.07$\pm$0.03 & -0.06$\pm$0.05 & -0.13$\pm$0.10 & 0.01$\pm$0.10 & 1 \\ 
        K2-265 & 5466$\pm$26 & 4.36$\pm$0.05 & 0.77$\pm$0.05 & 0.09$\pm$0.02 & 0.11$\pm$0.03 & 0.05$\pm$0.04 & 0.00$\pm$0.10 & 0.07$\pm$0.10  & 1 \\ 
        K2-291 & 5541$\pm$24 & 4.39$\pm$0.06 & 0.97$\pm$0.04 & 0.09$\pm$0.02 & 0.10$\pm$0.03 & 0.04$\pm$0.04 & 0.00$\pm$0.10 & 0.07$\pm$0.10  & 1 \\ 
        K2-38 & 5731$\pm$66 & 4.38$\pm$0.11 & 0.98$\pm$0.03 & 0.26$\pm$0.05 & 0.24$\pm$0.05 & 0.27$\pm$0.06 & 0.21$\pm$0.10 & 0.25$\pm$0.10 & 1 \\ 
        Kepler-10 & 5669$\pm$16 & 4.33$\pm$0.03 & 0.92$\pm$0.02 & -0.14$\pm$0.01 & -0.01$\pm$0.03 & -0.10$\pm$0.03 & -0.10$\pm$0.10 & 0.01$\pm$0.10 & 1 \\ 
        Kepler-107 & 5958$\pm$37 & 4.21$\pm$0.05 & 1.25$\pm$0.04 & 0.42$\pm$0.03 & 0.41$\pm$0.11 & 0.37$\pm$0.04 & 0.32$\pm$0.10 & 0.42$\pm$0.10 & 1 \\ 
        Kepler-20 & 5502$\pm$35 & 4.40$\pm$0.06 & 0.83$\pm$0.06 & 0.05$\pm$0.03 & 0.07$\pm$0.04 & 0.06$\pm$0.03 & 0.00$\pm$0.10 & 0.05$\pm$0.10 & 1 \\ 
        Kepler-21 & 6331$\pm$30 & 4.20$\pm$0.03 & 1.45$\pm$0.03 & 0.04$\pm$0.02 & -0.03$\pm$0.05 & -0.01$\pm$0.03 & -0.09$\pm$0.10 & 0.03$\pm$0.10 & 2 \\ 
        Kepler-78 & 4993$\pm$72 & 4.43$\pm$0.20 & 0.97$\pm$0.16 & -0.14$\pm$0.04 & -0.15$\pm$0.10 & -0.09$\pm$0.07 & -0.16$\pm$0.10 & 0.01$\pm$0.10 & 1 \\ 
        Kepler-93 & 5620$\pm$17 & 4.41$\pm$0.03 & 0.72$\pm$0.03 & -0.14$\pm$0.01 & -0.14$\pm$0.04 & -0.16$\pm$0.02 & -0.19$\pm$0.10 & 0.01$\pm$0.10 & 1 \\ 
        TOI-431 & 4850$\pm$75 & 4.60$\pm$0.06 & 0.80$\pm$0.10 & 0.20$\pm$0.05 & 0.10$\pm$0.07 & 0.11$\pm$0.13 & 0.06$\pm$0.10 & 0.10$\pm$0.10 & 6\\ 
        TOI-561 & 5314$\pm$20 & 4.37$\pm$0.04 & 0.54$\pm$0.05 & -0.39$\pm$0.02 & -0.18$\pm$0.04 & -0.27$\pm$0.03 & -0.28$\pm$0.10 & 0.01$\pm$0.10 & 1 \\ 
        TOI-733 & 5652$\pm$16 & 4.42$\pm$0.04 & 0.90$\pm$0.03 & -0.04$\pm$0.01 & -0.02$\pm$0.04 & -0.05$\pm$0.01 & -0.10$\pm$0.10 & 0.01$\pm$0.10 & 2 \\ 
        WASP-47 & 5559$\pm$52 & 4.32$\pm$0.12 & 1.09$\pm$0.07 & 0.40$\pm$0.04 & 0.46$\pm$0.06 & 0.43$\pm$0.05 & 0.35$\pm$0.10 & 0.42$\pm$0.10 & 1 \\ \hline
    \end{tabular}
    \tablebib{(1)~\citet{Adibekyan-21}; (2)~\citet{Adibekyan-24}; (3)~\citet{Barros-22}; (4)~\citet{Damasso-23}; (5)~This work; (6)~\citet{Osborn-21}}
    \label{table:params_abundances}
\end{table*}

\begin{table*}[]
\caption{\label{tab:init_abundances} The ages and primordial abundances of the planet host stars.}
\centering
\small
\begin{tabular}{lllllll}
\hline\hline

        star &  [Fe/H]$_{init}$ (dex) & [Mg/H]$_{init}$ (dex) & [Si/H]$_{init}$ (dex) & [C/H]$_{init}$ (dex) & [O/H]$_{init}$ (dex) & Age (Gyr) \\ \hline
        55 Cnc & 0.37$\pm$0.07 & 0.46$\pm$0.08 & 0.39$\pm$0.07 & 0.32$\pm$0.11 & 0.37$\pm$0.11 & 7.53$_{-7.51}^{+4.31}$ \\ 
        CoRoT-7 & 0.10$\pm$0.03 & 0.10$\pm$0.04 & 0.14$\pm$0.08 & 0.06$\pm$0.10 & 0.12$\pm$0.10 & 13.19$_{-0.41}^{+0.36}$ \\ 
        EPIC 220674823 & 0.17$\pm$0.04 & 0.14$\pm$0.06 & 0.12$\pm$0.04 & 0.07$\pm$0.10 & 0.12$\pm$0.10 & 9.97$_{-2.51}^{+1.76}$ \\ 
        EPIC 249893012 & 0.26$\pm$0.04 & 0.30$\pm$0.06 & 0.27$\pm$0.04 & 0.18$\pm$0.10 & 0.24$\pm$0.10 & 10.77$_{-1.28}^{+1.18}$ \\ 
        HD 136352 & -0.26$\pm$0.02 & -0.06$\pm$0.03 & -0.15$\pm$0.04 & -0.14$\pm$0.10 & 0.09$\pm$0.10 & 12.99$_{-0.54}^{+0.41}$ \\ 
        HD 137496 & 0.03$\pm$0.06 & 0.07$\pm$0.05 & 0.03$\pm$0.06 & 0.01$\pm$0.11 & 0.08$\pm$0.11 & 7.41$_{-1.61}^{+1.55}$ \\ 
        HD 15337 & 0.08$\pm$0.04 & 0.14$\pm$0.05 & 0.11$\pm$0.06 & 0.06$\pm$0.10 & 0.12$\pm$0.10 & 8.94$_{-7.99}^{+3.88}$ \\ 
        HD 20329 & -0.02$\pm$0.03 & 0.16$\pm$0.05 & 0.11$\pm$0.05 & 0.07$\pm$0.10 & 0.13$\pm$0.10 & 12.95$_{-0.37}^{+0.32}$ \\ 
        HD 213885 & 0.05$\pm$0.02 & 0.04$\pm$0.04 & 0.04$\pm$0.03 & -0.01$\pm$0.10 & 0.07$\pm$0.10 & 4.62$_{-1.09}^{+1.03}$ \\ 
        HD 219134 & 0.01$\pm$0.04 & 0.01$\pm$0.07 & 0.04$\pm$0.06 & -0.05$\pm$0.10 & 0.06$\pm$0.10 & 6.54$_{-5.96}^{+5.73}$ \\ 
        HD 23472 & -0.14$\pm$0.06 & -0.13$\pm$0.10 & -0.12$\pm$0.09 & -0.19$\pm$0.11 & 0.07$\pm$0.11 & 13.20$_{-0.32}^{+0.26}$ \\ 
        HD 3167 & 0.11$\pm$0.04 & 0.16$\pm$0.05 & 0.09$\pm$0.05 & 0.05$\pm$0.10 & 0.11$\pm$0.10 & 10.56$_{-8.01}^{+2.57}$ \\ 
        HD 80653 & 0.36$\pm$0.03 & 0.33$\pm$0.03 & 0.36$\pm$0.03 & 0.29$\pm$0.10 & 0.35$\pm$0.10 & 2.49$_{-1.57}^{+1.00}$ \\ 
        HIP 29442 & 0.32$\pm$0.06 & 0.34$\pm$0.05 & 0.29$\pm$0.05 & 0.25$\pm$0.10 & 0.32$\pm$0.10 & 13.28$_{-0.36}^{+0.32}$ \\ 
        K2-111 & -0.41$\pm$0.02 & -0.13$\pm$0.05 & -0.23$\pm$0.04 & -0.24$\pm$0.10 & 0.08$\pm$0.10 & 9.69$_{-6.81}^{+3.10}$ \\ 
        K2-131 & -0.05$\pm$0.03 & -0.07$\pm$0.05 & -0.05$\pm$0.04 & -0.09$\pm$0.10 & 0.08$\pm$0.10 & 12.31$_{-0.63}^{+0.58}$ \\ 
        K2-229 & 0.00$\pm$0.03 & -0.01$\pm$0.03 & 0.00$\pm$0.05 & -0.07$\pm$0.10 & 0.07$\pm$0.10 & 10.00$_{-9.37}^{+3.02}$ \\ 
        K2-265 & 0.16$\pm$0.03 & 0.17$\pm$0.03 & 0.11$\pm$0.05 & 0.07$\pm$0.10 & 0.13$\pm$0.10 & 10.24$_{-2.94}^{+2.04}$ \\ 
        K2-291 & 0.15$\pm$0.03 & 0.15$\pm$0.03 & 0.09$\pm$0.04 & 0.06$\pm$0.10 & 0.12$\pm$0.10 & 7.67$_{-3.43}^{+2.54}$ \\ 
        K2-38 & 0.29$\pm$0.08 & 0.27$\pm$0.08 & 0.30$\pm$0.08 & 0.25$\pm$0.12 & 0.28$\pm$0.12 & 4.10$_{-2.82}^{+2.20}$ \\ 
        Kepler-10 & -0.06$\pm$0.02 & 0.07$\pm$0.03 & -0.02$\pm$0.03 & -0.02$\pm$0.10 & 0.09$\pm$0.10 & 10.77$_{-1.69}^{+1.32}$ \\ 
        Kepler-107 & 0.45$\pm$0.04 & 0.44$\pm$0.11 & 0.40$\pm$0.05 & 0.35$\pm$0.10 & 0.45$\pm$0.10 & 3.22$_{-3.19}^{+0.51}$ \\ 
        Kepler-20 & 0.11$\pm$0.04 & 0.13$\pm$0.05 & 0.12$\pm$0.04 & 0.06$\pm$0.10 & 0.11$\pm$0.10 & 8.41$_{-3.68}^{+2.78}$ \\ 
        Kepler-21 & 0.09$\pm$0.03 & 0.01$\pm$0.05 & 0.04$\pm$0.03 & -0.04$\pm$0.10 & 0.08$\pm$0.10 & 2.86$_{-0.15}^{+0.16}$ \\ 
        Kepler-78 & -0.10$\pm$0.06 & -0.11$\pm$0.11 & -0.05$\pm$0.08 & -0.12$\pm$0.11 & 0.05$\pm$0.11 & 7.18$_{-5.40}^{+5.04}$ \\ 
        Kepler-93 & -0.07$\pm$0.02 & -0.08$\pm$0.05 & -0.10$\pm$0.03 & -0.12$\pm$0.10 & 0.07$\pm$0.10 & 9.09$_{-2.36}^{+2.02}$ \\ 
        TOI-431 & 0.24$\pm$0.06 & 0.13$\pm$0.07 & 0.14$\pm$0.13 & 0.09$\pm$0.10 & 0.13$\pm$0.10 & 5.68$_{-0.80}^{+0.95}$ \\ 
        TOI-561 & -0.33$\pm$0.04 & -0.11$\pm$0.05 & -0.20$\pm$0.04 & -0.21$\pm$0.10 & 0.08$\pm$0.10 & 12.93$_{-2.52}^{+0.68}$ \\ 
        TOI-733 & 0.02$\pm$0.02 & 0.03$\pm$0.04 & 0.00$\pm$0.02 & -0.04$\pm$0.10 & 0.06$\pm$0.10 & 7.39$_{-0.53}^{+0.60}$ \\ 
        WASP-47 & 0.44$\pm$0.06 & 0.50$\pm$0.07 & 0.47$\pm$0.06 & 0.39$\pm$0.11 & 0.46$\pm$0.11 & 5.41$_{-3.45}^{+2.45}$ \\ \hline
    \end{tabular}
\end{table*}

\begin{table*}[]
    \caption{\label{tab:building_blocks} The present-day and primordial compositions of planet building blocks.}
    \centering
    \small
    \begin{tabular}{lllll}
    \hline\hline
        star & $f_{\mathrm{iron}}^{\mathrm{star}}$ (\%) & $f_{\mathrm{water}}^{\mathrm{star}}$ (\%) & $f_{\mathrm{iron,prim}}^{\mathrm{star}}$ (\%) & $f_{\mathrm{water,prim}}^{\mathrm{star}}$ (\%) \\
        \hline
        55 Cnc & 29.3$\pm$2.7 & 55.6$\pm$7.6 & 29.3$\pm$2.7 & 55.5$\pm$8.0 \\
        EPIC 249893012 & 31.4$\pm$1.5 & 55.5$\pm$7.5 & 31.3$\pm$1.5 & 55.1$\pm$7.5 \\
        HD 137496 & 31.4$\pm$2.1 & 60.7$\pm$6.7 & 31.5$\pm$2.2 & 60.5$\pm$6.9 \\
        HD 213885 & 33.1$\pm$1.3 & 60.4$\pm$6.9 & 32.9$\pm$1.3 & 59.7$\pm$6.9 \\
        HD 219134 & 31.4$\pm$2.4 & 60.3$\pm$7.2 & 31.3$\pm$2.4 & 60.2$\pm$7.2 \\
        HD 3167 & 31.3$\pm$1.8 & 57.8$\pm$7.2 & 31.4$\pm$1.9 & 57.6$\pm$7.2 \\
        HD 80653 & 33.5$\pm$1.3 & 57.9$\pm$7.1 & 33.5$\pm$1.4 & 57.8$\pm$7.3 \\
        EPIC 220674823 & 34.3$\pm$2.0 & 56.5$\pm$7.4 & 34.3$\pm$2.0 & 56.0$\pm$7.3 \\
        K2-111 & 22.5$\pm$1.1 & 76.6$\pm$4.8 & 22.6$\pm$1.1 & 76.4$\pm$4.6 \\
        K2-229 & 32.5$\pm$1.8 & 63.0$\pm$6.7 & 32.5$\pm$1.8 & 62.8$\pm$6.5 \\
        K2-265 & 32.8$\pm$1.7 & 57.3$\pm$7.1 & 32.8$\pm$1.6 & 57.6$\pm$7.3 \\
        K2-291 & 33.3$\pm$1.5 & 58.2$\pm$7.1 & 33.3$\pm$1.5 & 58.1$\pm$6.6 \\
        K2-38 & 32.3$\pm$3.2 & 57.7$\pm$7.2 & 32.4$\pm$3.2 & 57.4$\pm$7.6 \\
        Kepler-10 & 28.4$\pm$1.2 & 64.0$\pm$6.5 & 28.4$\pm$1.1 & 64.1$\pm$6.6 \\
        Kepler-107 & 33.9$\pm$3.2 & 59.5$\pm$7.3 & 33.8$\pm$3.2 & 59.2$\pm$7.4 \\
        Kepler-20 & 31.5$\pm$1.8 & 57.1$\pm$7.3 & 31.5$\pm$1.8 & 57.5$\pm$7.2 \\
        Kepler-78 & 30.8$\pm$3.4 & 66.4$\pm$6.5 & 31.0$\pm$3.4 & 66.4$\pm$6.5 \\
        Kepler-93 & 32.8$\pm$1.3 & 68.1$\pm$5.6 & 32.8$\pm$1.3 & 68.1$\pm$5.7 \\
        HD 15337 & 30.0$\pm$2.0 & 58.6$\pm$7.1 & 30.1$\pm$2.1 & 58.8$\pm$7.0 \\
        TOI-561 & 24.5$\pm$1.5 & 74.5$\pm$5.0 & 24.6$\pm$1.4 & 74.4$\pm$5.3 \\
        WASP-47 & 29.8$\pm$2.7 & 57.2$\pm$7.4 & 29.9$\pm$2.6 & 56.8$\pm$7.8 \\
        HD 20329 & 24.7$\pm$1.5 & 60.6$\pm$7.0 & 24.7$\pm$1.5 & 60.6$\pm$7.2 \\
        Kepler-21 & 34.9$\pm$1.7 & 59.7$\pm$6.9 & 35.0$\pm$1.7 & 60.2$\pm$7.1 \\
        CoRoT-7 & 30.7$\pm$2.5 & 58.5$\pm$7.5 & 30.5$\pm$2.6 & 57.8$\pm$7.7 \\
        HIP 29442 & 32.8$\pm$2.8 & 58.7$\pm$7.1 & 32.8$\pm$2.9 & 58.5$\pm$7.0 \\
        TOI-733 & 32.3$\pm$1.0 & 61.4$\pm$6.6 & 32.3$\pm$1.0 & 61.0$\pm$6.8 \\
        TOI-431 & 37.1$\pm$4.9 & 55.3$\pm$8.4 & 36.8$\pm$5.2 & 55.1$\pm$8.4 \\
        K2-131 & 32.6$\pm$1.6 & 66.6$\pm$6.1 & 32.8$\pm$1.6 & 67.3$\pm$6.2 \\
        HD 23472 & 31.3$\pm$3.9 & 70.2$\pm$6.1 & 31.7$\pm$3.8 & 70.4$\pm$6.2 \\
        HD 136352 & 25.0$\pm$1.1 & 72.4$\pm$5.3 & 25.1$\pm$1.1 & 72.6$\pm$5.4 \\ \hline
    \end{tabular}
\end{table*}

\begin{table*}[]
    \caption{\label{tab:planet_composition} The compositions of planets.}
    \centering
    \small
    \begin{tabular}{llllll}
    \hline\hline
        Planet & $f_{\mathrm{mantle}}$ (\%) & $f_{\mathrm{Fe}}$ (\%) & $f_{\mathrm{water}}$ (\%) & $f_{\mathrm{iron}}^{\mathrm{planet}}$ & (Fe/Mg)$_{\mathrm{planet}}$ \\ \hline
        55 Cnc e & 65.4$_{-20.8}^{+11.1}$ & 32.3$_{-9.8}^{+16.2}$ & 2.3$_{-1.6}^{+4.7}$ & 33.1$_{-9.7}^{+10.3}$ & 0.9$_{-0.4}^{+1.1}$ \\
        EPIC 249893012 b & 88.2$_{-7.5}^{+7.8}$ & 11.8$_{-7.8}^{+7.5}$ & 0.0$_{-0.0}^{+0.0}$ & 11.1$_{-6.8}^{+6.5}$ & 0.2$_{-0.2}^{+0.2}$ \\
        HD 137496 b & 23.6$_{-11.0}^{+12.2}$ & 76.1$_{-11.9}^{+11.0}$ & 0.1$_{-0.1}^{+0.6}$ & 77.0$_{-9.5}^{+9.0}$ & 5.7$_{-2.5}^{+6.6}$ \\
        HD 213885 b & 51.8$_{-16.5}^{+17.7}$ & 47.7$_{-17.3}^{+14.7}$ & 0.2$_{-0.1}^{+2.2}$ & 47.6$_{-11.9}^{+12.6}$ & 1.6$_{-0.9}^{+1.5}$ \\
        HD 219134 b & 59.9$_{-23.8}^{+14.8}$ & 37.9$_{-13.7}^{+19.2}$ & 1.9$_{-1.4}^{+5.7}$ & 38.6$_{-13.6}^{+13.9}$ & 1.1$_{-0.6}^{+1.7}$ \\
        HD 219134 c & 37.1$_{-19.0}^{+16.6}$ & 56.8$_{-13.8}^{+15.2}$ & 5.8$_{-3.3}^{+4.6}$ & 62.1$_{-12.4}^{+11.4}$ & 2.7$_{-1.3}^{+4.3}$ \\
        HD 3167 b & 79.1$_{-11.2}^{+10.9}$ & 20.9$_{-10.9}^{+11.1}$ & 0.0$_{-0.0}^{+0.2}$ & 19.7$_{-9.4}^{+8.9}$ & 0.5$_{-0.3}^{+0.4}$ \\
        HD 80653 b & 56.6$_{-21.5}^{+20.6}$ & 42.8$_{-20.1}^{+20.6}$ & 0.1$_{-0.1}^{+1.3}$ & 42.2$_{-15.2}^{+16.1}$ & 1.3$_{-0.8}^{+1.9}$ \\
        EPIC 220674823 b & 50.6$_{-13.7}^{+17.3}$ & 49.0$_{-16.9}^{+12.3}$ & 0.3$_{-0.3}^{+1.5}$ & 49.5$_{-11.6}^{+11.6}$ & 1.7$_{-0.9}^{+1.2}$ \\
        K2-111 b & 73.0$_{-37.6}^{+18.7}$ & 24.9$_{-17.2}^{+26.4}$ & 1.2$_{-1.2}^{+12.5}$ & 23.4$_{-17.4}^{+19.6}$ & 0.6$_{-0.5}^{+1.9}$ \\
        K2-229 b & 29.7$_{-12.4}^{+16.1}$ & 70.2$_{-16.1}^{+12.4}$ & 0.0$_{-0.0}^{+0.0}$ & 70.4$_{-10.5}^{+10.2}$ & 4.2$_{-2.1}^{+4.3}$ \\
        K2-265 b & 58.2$_{-26.8}^{+21.7}$ & 39.6$_{-20.0}^{+22.9}$ & 0.8$_{-0.8}^{+7.3}$ & 41.0$_{-17.9}^{+17.7}$ & 1.2$_{-0.8}^{+2.3}$ \\
        K2-291 b & 52.2$_{-27.3}^{+27.3}$ & 46.9$_{-26.4}^{+24.9}$ & 0.2$_{-0.2}^{+2.7}$ & 46.7$_{-20.3}^{+19.6}$ & 1.6$_{-1.1}^{+3.5}$ \\
        K2-38 b & 48.9$_{-22.6}^{+23.7}$ & 49.9$_{-23.1}^{+20.2}$ & 0.3$_{-0.2}^{+3.7}$ & 51.7$_{-16.6}^{+16.1}$ & 1.8$_{-1.1}^{+2.9}$ \\
        Kepler-10 b & 71.9$_{-21.6}^{+13.1}$ & 28.0$_{-13.0}^{+20.7}$ & 0.0$_{-0.0}^{+0.9}$ & 25.0$_{-10.8}^{+15.1}$ & 0.7$_{-0.4}^{+1.0}$ \\
        Kepler-107 c & 19.0$_{-7.9}^{+12.1}$ & 79.9$_{-11.7}^{+8.0}$ & 0.3$_{-0.2}^{+1.2}$ & 81.4$_{-8.0}^{+8.3}$ & 7.5$_{-3.5}^{+6.6}$ \\
        Kepler-20 b & 52.0$_{-16.2}^{+15.1}$ & 47.5$_{-14.9}^{+14.5}$ & 0.2$_{-0.1}^{+2.0}$ & 47.2$_{-10.6}^{+10.5}$ & 1.6$_{-0.8}^{+1.4}$ \\
        Kepler-78 b & 52.0$_{-22.8}^{+22.5}$ & 48.0$_{-22.4}^{+22.6}$ & 0.0$_{-0.0}^{+0.2}$ & 47.4$_{-15.4}^{+17.4}$ & 1.6$_{-1.0}^{+2.7}$ \\
        Kepler-93 b & 42.6$_{-9.8}^{+10.6}$ & 56.3$_{-10.2}^{+9.1}$ & 1.0$_{-0.6}^{+1.2}$ & 57.5$_{-7.5}^{+7.5}$ & 2.3$_{-0.8}^{+1.2}$ \\
        HD 15337 b & 65.6$_{-24.1}^{+15.2}$ & 33.6$_{-14.4}^{+21.9}$ & 0.1$_{-0.1}^{+4.2}$ & 32.6$_{-12.8}^{+14.8}$ & 0.9$_{-0.5}^{+1.4}$ \\
        TOI-561 b & 75.7$_{-18.4}^{+13.6}$ & 24.3$_{-13.5}^{+18.1}$ & 0.0$_{-0.0}^{+0.2}$ & 22.6$_{-12.8}^{+13.3}$ & 0.6$_{-0.4}^{+0.7}$ \\
        WASP-47 e & 71.3$_{-22.7}^{+10.8}$ & 27.8$_{-10.3}^{+18.8}$ & 0.8$_{-0.7}^{+4.0}$ & 27.8$_{-11.4}^{+12.3}$ & 0.7$_{-0.3}^{+1.0}$ \\
        HD 20329 b & 48.5$_{-22.2}^{+21.5}$ & 50.4$_{-20.9}^{+17.7}$ & 0.8$_{-0.7}^{+5.1}$ & 51.9$_{-16.2}^{+14.6}$ & 1.8$_{-1.1}^{+2.7}$ \\
        Kepler-21 b & 38.9$_{-10.0}^{+12.1}$ & 60.2$_{-11.6}^{+9.7}$ & 0.5$_{-0.4}^{+1.2}$ & 60.0$_{-8.1}^{+8.2}$ & 2.7$_{-1.0}^{+1.5}$ \\
        CoRoT-7 b & 36.8$_{-14.8}^{+16.9}$ & 62.2$_{-16.3}^{+14.2}$ & 0.3$_{-0.2}^{+1.8}$ & 63.3$_{-12.1}^{+11.5}$ & 3.0$_{-1.5}^{+3.2}$ \\
        HIP 29442 c & 69.8$_{-24.3}^{+15.3}$ & 29.3$_{-14.7}^{+22.3}$ & 0.5$_{-0.5}^{+2.4}$ & 27.4$_{-13.1}^{+15.6}$ & 0.7$_{-0.4}^{+1.3}$ \\
        TOI-733 b & 52.0$_{-27.1}^{+29.5}$ & 29.5$_{-18.8}^{+20.4}$ & 17.9$_{-10.6}^{+8.9}$ & 29.1$_{-14.2}^{+28.1}$ & 1.0$_{-0.8}^{+2.5}$ \\
        TOI-431 b & 37.6$_{-11.9}^{+13.0}$ & 62.3$_{-13.0}^{+11.8}$ & 0.0$_{-0.0}^{+0.2}$ & 63.0$_{-9.0}^{+9.5}$ & 2.9$_{-1.2}^{+2.2}$ \\
        K2-131 b & 71.7$_{-33.2}^{+12.2}$ & 28.3$_{-12.2}^{+32.1}$ & 0.0$_{-0.0}^{+0.5}$ & 22.0$_{-12.3}^{+22.8}$ & 0.7$_{-0.4}^{+2.1}$ \\
        HD 23472 b & 83.7$_{-19.7}^{+10.8}$ & 14.7$_{-9.8}^{+14.8}$ & 0.1$_{-0.1}^{+7.9}$ & 14.1$_{-11.1}^{+11.6}$ & 0.3$_{-0.2}^{+0.5}$ \\
        HD 23472 c & 61.2$_{-35.0}^{+26.3}$ & 20.8$_{-14.9}^{+20.0}$ & 20.9$_{-20.8}^{+14.4}$ & 28.9$_{-21.8}^{+18.5}$ & 0.6$_{-0.5}^{+2.1}$ \\
        HD 136352 b & 85.5$_{-10.6}^{+9.0}$ & 14.3$_{-8.8}^{+10.4}$ & 0.0$_{-0.0}^{+0.6}$ & 14.0$_{-8.1}^{+8.1}$ & 0.3$_{-0.2}^{+0.3}$ \\ \hline
    \end{tabular}
    \tablefoot{$f_{\mathrm{mantle}}$,  $f_{\mathrm{Fe}}$ (\%), and  $f_{\mathrm{water}}$  are the mantle, iron, and water mass fractions. $f_{\mathrm{iron}}^{\mathrm{planet}}$ is the ratio of $f_{\mathrm{Fe}}$ / ($f_{\mathrm{Fe}}$ + $f_{\mathrm{mantle}}$).
}
\end{table*}

\section{The star--planet compositional relationship} \label{apdx:relation}

\citet{Brinkman-24} recently suggested that both the slope and the significance of the relationship between the iron mass fraction of stars and their rocky planets depend on the fitting algorithm used. In this section, we apply several algorithms to test the robustness of this relationship.

The OLS regression used in the manuscript does not account for the uncertainties in $f_{\mathrm{iron}}^{\mathrm{planet}}$ and $f_{\mathrm{iron,prim}}^{\mathrm{star}}$. To address this, we performed a Monte Carlo (MC) test. Assuming Gaussian distributions for the uncertainties in these parameters, centered on their respective values, we randomly sampled values for $f_{\mathrm{iron}}^{\mathrm{planet}}$ and $f_{\mathrm{iron,prim}}^{\mathrm{star}}$ neglecting the cases when the values are negative. After 10000 realizations, we obtained a median slope of 1.65$\pm$0.78, with a corresponding p-value (P(t-stat)) of 0.043.

As a second method, we applied the Weighted Least Squares (WLS) regression as implemented in \texttt{statsmodels} package, using the inverse square of the uncertainties in $f_{\mathrm{iron}}^{\mathrm{planet}}$ as weights. This yielded a slope of 3.85$\pm$1.10 with a p-value of 0.002. We then repeated the previously described MC test for the WLS case. After 10,000 realizations, we obtained a median slope of 2.43$\pm$0.89, with a p-value of 0.01.

For the third method, we used Orthogonal Distance Regression (ODR), implemented in \texttt{SCIPY}, which accounts for uncertainties in both parameters during fitting. ODR resulted in a slope of 5.98$\pm$1.08, with a p-value of about $10^{-5}$. Repeating the MC test for ODR over 10,000 realizations, we obtained a median slope of 5.85$\pm$1.07, leading to a p-value of $\sim10^{-5}$.

Our fourth method was Total Least Squares (TLS), a generalization of ODR that also considers observational errors in both variables. For the log-likelihood calculation, we used \texttt{astroML} package which follows the prescription of \citet{Hogg-10}, and then used the \texttt{scipy.optimize} package to maximize it and find the best-fit parameters. After performing the MC test for the TLS method with 10,000 realizations, we obtained a median slope of 24.9$\pm$0.9, with a p-value below $10^{-22}$.

Finally, we applied \texttt{PyStan}, a Python interface to the Bayesian inference package \texttt{Stan} \citep{Carpenter-17} which allowed us to account for uncertainties in both parameters within a Bayesian framework. We used wide priors for the slope, intercept, and sigma (the scale of the noise term), modeled as normal distributions centered at zero with a standard deviation of 100. From the posterior distributions of the slopes, we obtained a median value of 3.06, with a 95\% credible interval of [0.8, 5.0]. Notably, less than 4\% of the posterior slopes were smaller than 1.

\begin{figure}
\begin{center}
\includegraphics[width=0.9\linewidth]{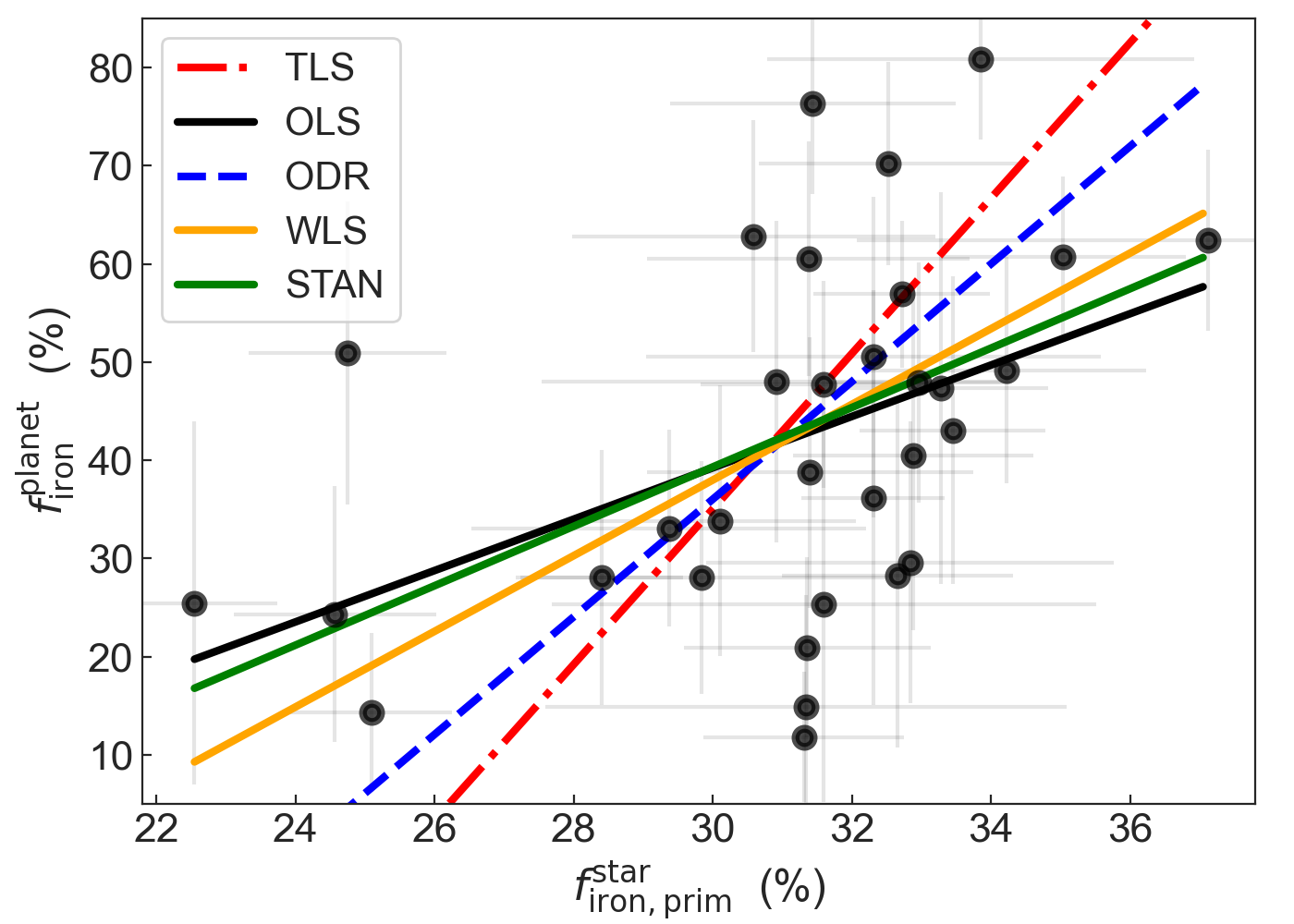}
\end{center}
\vspace{-0.4cm}
\caption{Iron-to-silicate mass fractions of planets and their host stars, with different lines representing the results of the linear fits discussed in the manuscript.}
\label{fig:all_slopes}
\end{figure}

Overall, these tests demonstrate a statistically significant relationship between $f_{\mathrm{iron}}^{\mathrm{planet}}$ and $f_{\mathrm{iron,prim}}^{\mathrm{star}}$, with a slope greater than one. In Fig.\,\ref{fig:all_slopes} we show the results of the aforementioned linear fits.

\end{appendix}

\end{document}